\documentclass[aps,reprint,
superscriptaddress,floatfix,longbibliography,footinbib]{revtex4-1}
\usepackage{mathtools}
\usepackage{amsfonts}
\usepackage{amsmath}
\usepackage{amssymb}
\usepackage{physics}
\usepackage{float}
\usepackage{bbold}
\usepackage[version=4]{mhchem}
\usepackage{hyperref}
\usepackage{xcolor}
\usepackage{lipsum}
\usepackage{xspace} 
\usepackage[rightcaption]{sidecap}
\usepackage{nicematrix}
\usepackage{diagbox}

\usepackage{tabularray}
\UseTblrLibrary{booktabs}

\hypersetup{
   colorlinks=true,
   linkcolor=blue,
   filecolor=blue, 
   citecolor=blue,
   urlcolor=blue,
}
\usepackage[doipre={DOI:~}]{uri}

\newcommand{\eigmax}{{\lambda_{\mathrm{max}}}}

\newcommand{\ex}[1]{{\langle #1 \rangle}}

\newcommand{\mbf}{\mathbf}

\begin{document}

\title{A multimode cavity QED Ising spin glass}

\author{Brendan P.~Marsh}
\altaffiliation[B.~M. and D.~A.~S.~contributed equally to this work.]{}
\affiliation{Department of Applied Physics, Stanford University, Stanford, CA 94305, USA}
\affiliation{E.~L.~Ginzton Laboratory, Stanford University, Stanford, CA 94305, USA}
\author{David Atri Schuller}
\altaffiliation[B.~M. and D.~A.~S.~contributed equally to this work.]{}
\affiliation{Department of Applied Physics, Stanford University, Stanford, CA 94305, USA}
\affiliation{E.~L.~Ginzton Laboratory, Stanford University, Stanford, CA 94305, USA}
\author{Yunpeng Ji}
\affiliation{Department of Applied Physics, Stanford University, Stanford, CA 94305, USA}
\affiliation{E.~L.~Ginzton Laboratory, Stanford University, Stanford, CA 94305, USA}
\affiliation{Department of Physics, Stanford University, Stanford, CA 94305, USA}
\author{Henry S.~Hunt}
\affiliation{E.~L.~Ginzton Laboratory, Stanford University, Stanford, CA 94305, USA}
\affiliation{Department of Physics, Stanford University, Stanford, CA 94305, USA}
\author{\\Giulia Z.~Socolof}
\affiliation{Department of Applied Physics, Stanford University, Stanford, CA 94305, USA}
\affiliation{E.~L.~Ginzton Laboratory, Stanford University, Stanford, CA 94305, USA}
\author{Deven P.~Bowman}
\affiliation{E.~L.~Ginzton Laboratory, Stanford University, Stanford, CA 94305, USA}
\affiliation{Department of Physics, Stanford University, Stanford, CA 94305, USA}
\author{Jonathan Keeling} 
\affiliation{SUPA, School of Physics and Astronomy, University of St. Andrews, St. Andrews KY16 9SS, United Kingdom}
\author{Benjamin L.~Lev}
\affiliation{Department of Applied Physics, Stanford University, Stanford, CA 94305, USA}
\affiliation{E.~L.~Ginzton Laboratory, Stanford University, Stanford, CA 94305, USA}
\affiliation{Department of Physics, Stanford University, Stanford, CA 94305, USA}

\date{May 28, 2025}

\begin{abstract}

We realize a driven-dissipative Ising spin glass using cavity QED in a novel ``4/7" multimode  geometry. Gases of ultracold atoms trapped within the cavity by optical tweezers serve as effective spins. 
They are coupled via randomly signed, all-to-all Ising cavity-mediated interactions. Networks of up to $n=25$ spins are holographically imaged via cavity emission. The system is driven through a frustrated transverse-field Ising transition, and we show that the entropy of the spin glass states depends on the rate at which the transition is crossed.   Despite being intrinsically nonequilibrium, the system exhibits phenomena associated with Parisi's theory of equilibrium spin glasses, namely replica symmetry breaking (RSB) and ultrametric structure.  For system sizes up to $n=16$, we measure the Parisi function $q(x)$, Edwards-Anderson overlap $q_\text{EA}$, and ultrametricity $K$-correlator; all indicate a deeply ordered spin glass under RSB. The system can serve as an associative memory and enable aging and rejuvenation studies in driven-dissipative spin glasses at the microscopic level.

\end{abstract}

\maketitle

Ising spin glasses are frustrated, disordered magnets that, colloquially speaking, possess spins randomly frozen up or down. This differs from disordered states like paramagnets whose spins are rapidly fluctuating or simply ordered states like ferromagnets whose spins are frozen, but nearly all up or down.  Real spin glasses are not actually frozen; rather, their spin configurations slowly evolve through metastable states and, it is believed, never equilibrate with their thermal environment~\cite{Stein2013sga}. But phases of matter can be out of equilibrium in other ways~\cite{Polkovnikov2011cnd,Sieberer2023uid,Fazio2024moq}, e.g., by being driven into steady states stabilized by dissipation.  This prompts the question:  Which features of glassiness persist in driven-dissipative systems?  More broadly, studying nonequilibrium systems might fundamentally expand our understanding of glass physics.

\begin{figure}[t!]
    \centering
    \includegraphics[width=\linewidth]{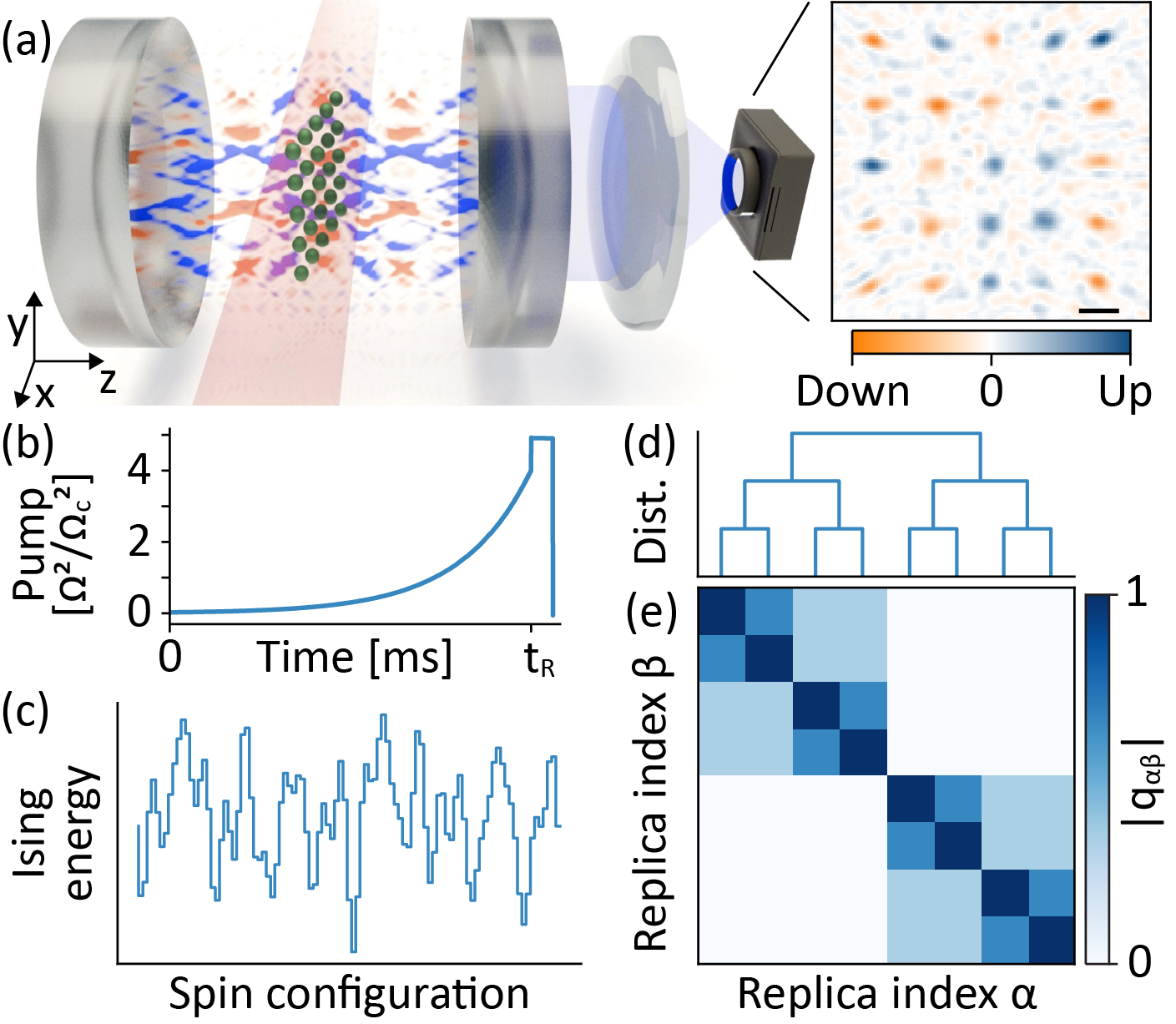}
    \caption{(a) Cartoon of the $4/7$ cavity QED apparatus. Ultracold atomic gases  (green) trapped in the cavity midplane serve as pseudospins. Light scattered by the atoms from a transverse pump (red) and into the cavity mediates spin interactions. A camera images cavity emission for spin state detection. A processed experimental image shows a $5{\times} 5$ array of spins; color indicates spin state. (b) The transverse pump strength is exponentially ramped to $4\times$ the critical power ${\propto}\Omega_c^2$ over a time $t_R$ before  quenching to a higher power for imaging. (c) 1D cartoon of a rugged, spin glass energy landscape. The actual space is high-dimensional. (d) Example dendrogram showing ultrametric structure arising from overlap distances $d_{\alpha\beta} = 1 - |q_{\alpha\beta}|$. (e) Example overlap matrix $q_{\alpha\beta}$ with  fractal ultrametric structure seen as block-diagonal correlations.  
    }
    \label{fig:Fig1}
\end{figure}

Models with all-to-all Ising coupling serve as paradigmatic  theories of spin glasses and recurrent neural networks like Hopfield's associative memory~\cite{Charbonneau2023sgt,Hopfield1982nna,Hopfield1986cwn,Bialek2024mba}.  However, these theories focus on thermal equilibrium.  We introduce a new platform that provides experimental access to driven-dissipative Ising spin glasses based on strongly coupled multimode cavity QED. This allows us to directly compare to the famous equilibrium Sherrington--Kirkpatrick (SK) model of Ising spin glass~\cite{Sherrington1975smo}.  To realize this coupling, we employ a resonator geometry different from previous studies, thereby adding a new capability to the multimode cavity QED toolbox~\cite{Vaidya2018tpa,Guo2019spa,Guo2019eab,Kroeze2023hcu,Kroeze2023rsb}. While the present work focuses on static properties, the system enables the study of questions regarding how dynamical properties like aging and rejuvenation differ from that of traditional Ising glasses~\cite{Charbonneau2023sgt} and how cavity QED dynamics  enhances associative memory capacity~\cite{Marsh2021eam,Marsh2025eam}.  

Parisi showed that the low-energy spin order of an equilibrium spin glass with all-to-all Ising coupling  exhibits an emergent structure among spin configurations called ultrametricity~\cite{Parisi1980top,Parisi1983opf}, defined below.   This structure arises through a peculiar broken symmetry: ``replicas," or identically prepared copies of the system, can independently evolve through a rugged energy landscape into different low-entropy states that are unrelated by any simple symmetry; Fig.~\ref{fig:Fig1}(c) sketches a glassy landscape showing the multitude of energy minima. Order manifests in the emergent structure of correlations among these replica states.  This theoretical result has had broad implications for the study of complex systems, including artificial neural networks~\cite{Stein2013sga,Parisi2023nlm}.

We previously reported the first direct observation of replica symmetry breaking (RSB) and nascent ultrametric structure in a driven-dissipative spin glass using a confocal multimode resonator~\cite{Kroeze2023rsb}. By ``direct," we mean at the level of experimentally measuring individual spin configurations.  RSB has also been observed with random lasers~\cite{ghofraniha2015eeo,gomes2016ool,pierangeli2017oor,Antenucci2015gpd,Niedda2023uco}, though not at the level of individual spin configurations.  A report subsequent to Ref.~\cite{Kroeze2023rsb} presents indirect measurements of ultrametric structure in those systems~\cite{Ghofraniha2024oou}. Superconducting circuits have been reported to operate near a spin glass transition~\cite{Harris2018pti,King2023qcd}.

The confocal cavity of Ref.~\cite{Kroeze2023rsb} yielded a vector ($XX{-}YY$) spin glass model of size $n=8$. To now create an all-to-all Ising ($XX$) coupling, we employ a new multimode cavity QED geometry---a ``4/7" resonator.  The network size has been tripled to 25, enabling the first experimental study of a spin glass versus network size.  The ability to image spin configurations allows us to measure the glass state entropy versus the rate at which we ramp through the transition. Moreover, it enables direct measurement of the Parisi function $q(x)$~\cite{Parisi1980aso}, the Edwards-Anderson order parameter $q_\text{EA}$~\cite{Edwards1975tos}, and the ultrametricity $K$-correlator~\cite{Katzgraber2009uac}.  Each is explained below and together quantify the full sense in which the driven-dissipative system is deeply ordered as an Ising spin glass.  Our measurements directly show RSB and ultrametricity at the individual spin-configuration level of an Ising glass, lending support to the observation that these are microscopically derivable properties of physical systems and not just abstractions of equilibrium theory models. 

\begin{figure*}[t!]
    \centering
    \includegraphics[width=\linewidth]{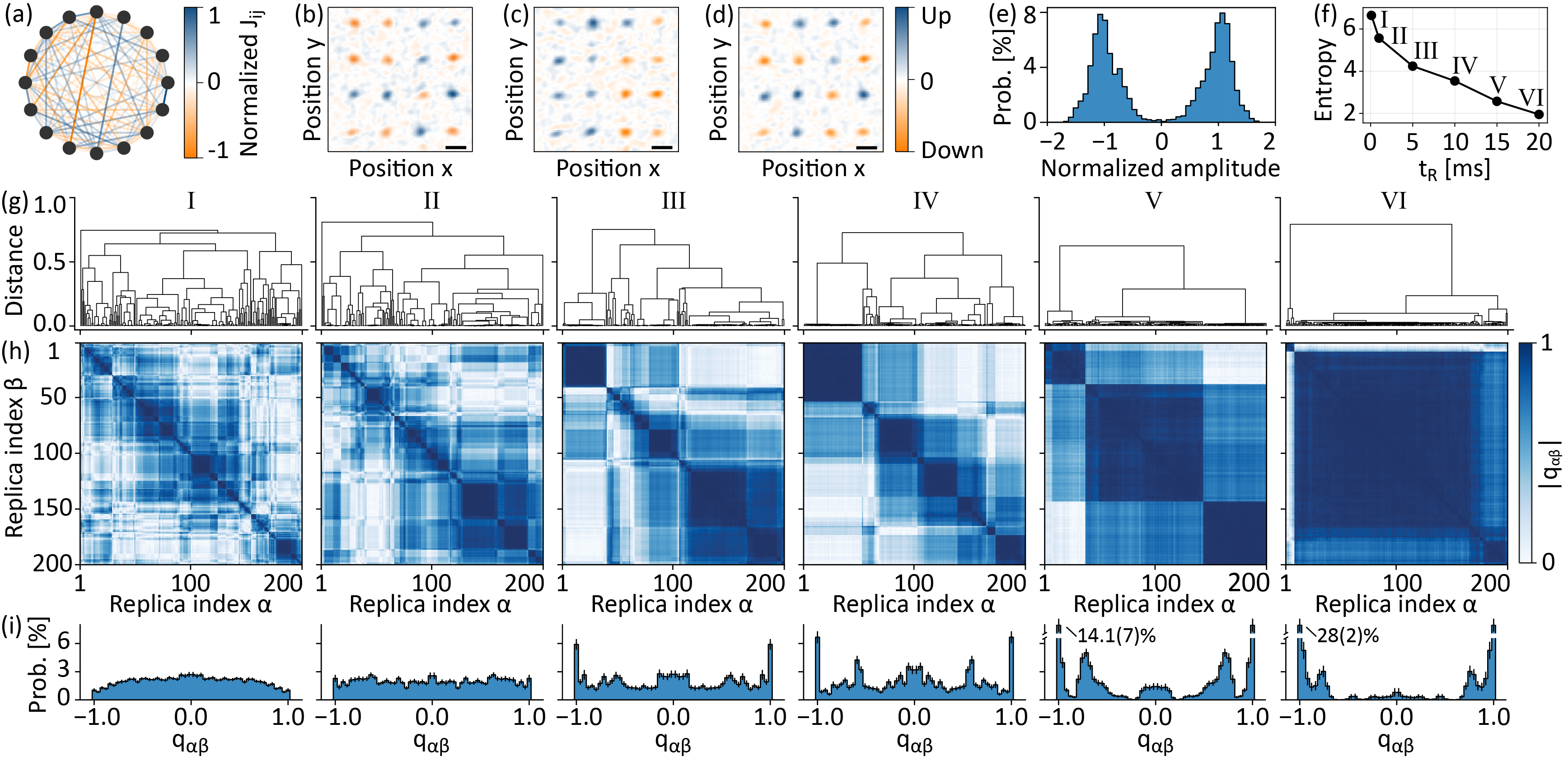}
    \caption{Study of spin configurations arising from the disorder realization $J_1$ of an $n= 16$ network. (a) $J_1$ connectivity diagram normalized to the absolute amplitude of the largest element. (b,c,d) Images of spin-state configurations of three replicas, normalized to the maximum spin amplitude in each. Scale bars are equal to the length of the Gaussian mode waist  $w_0=35$~$\mu$m.  (e) Spin amplitude distribution over all  spins in 200 replicas, normalized by the total RMS spin amplitude; Ramp time $t_R=5$~ms.   (f) Base-2 Shannon entropy of the binarized spin-state distribution versus $t_R$. Each point is derived from an ensemble of 200 replicas, and $t_R=\{0.1, 1, 5, 10, 15, 20\}$~ms from I to VI. (g) Hierarchical clustering of replicas by the overlap distance $d$. Columns I to VI correspond to ramp times in panel (f). (h) Overlap matrix $q_{\alpha\beta}$ and (i) overlap distribution.  Error bars are standard error both here and in figures below.}
    \label{fig:Fig2}
\end{figure*}

Figure~\ref{fig:Fig1} depicts the experimental system. The Fabry-P\'{e}rot cavity mirrors have an $R=1$-cm radius of curvature and are separated by a length $L\approx 1.22$~cm.  This geometry realizes a multimode cavity with a mode spectrum that differs from that of the confocal cavity previously employed~\cite{Vaidya2018tpa,Guo2019spa,Kroeze2023hcu}. A multimode degeneracy point occurs whenever the condition $L/R=2\sin^2(M\pi/2N)$ is satisfied for integers $M$ and $N$ forming an irreducible fraction $M/N$~\cite{Guo2019eab}\footnote{Imperfect degeneracy arises due to mirror aberrations~\cite{Kollar2015aac,Kroeze2023hcu}}.  The confocal cavity has $M/N=1/2$; here, we use an $M/N=4/7$ cavity; see Ref.~\cite{Supp} for details. Briefly, within each free spectral range there are $N$ mode families that resonate at distinct frequencies. Each is labeled by an integer $\eta$ and contains all Hermite-Gaussian modes $\Xi_{lm}$ that have $l+m\mod N=\eta$. $M$ determines the difference in longitudinal mode number $Q$ of the degenerate modes in each family.  An $\eta$ family supports those modes with $(Q,l+m)=(Q_0-Mi,\eta+Ni)$ for integers $i\geq0$, where $Q_0$ is the longitudinal mode number of the lowest-order transverse mode in the family. The longitudinal behavior varies cyclically with $Q$ as $\sin(kz),\cos(kz),-\sin(kz),-\cos(kz)$ up to a constant phase shift near the cavity midplane, where $k$ is the mode wavevector. As such, odd-$M$ cavities host families with modes of both $\cos{kz}$ and $\sin{kz}$ form, and superpositions of these modes yield a $U(1)$ phase degree of freedom of the cavity field. Cavity-mediated interactions in odd-M cavities therefore support a vector spin model~\cite{Kroeze2023rsb}. By contrast, even-$M$ cavities support either only $\cos{kz}$ or only $\sin{kz}$ modes in each family, resulting in a field that is either 0 or $\pi$ in phase. The resulting $\mathbb{Z}_2$ symmetry induces an Ising spin coupling.  Our demonstration of the utility of this unusual multimode cavity geometry illustrates the potentially wide range of spin models offered by $M/N$ cavities.

We produce atomic ensembles trapped at locations $\mbf{r}_i$ in the cavity midplane; each serves as a spin in the $n$-site network. To enhance light-matter coupling, each contains $N_A\approx$ $6 {\times} 10^4$ $^{87}$Rb atoms that are evaporated just below Bose degeneracy~\footnote{Matter-wave coherence plays no role in reported results.}. With a single-mode, single-atom coupling strength of $g_0=2\pi{\cdot}1.47$~MHz and field decay rate of $\kappa=2\pi{\cdot}140$~kHz, the cavity QED system lies within the strong coupling regime, even without multimode enhancement; see Refs.~\cite{Supp,Kroeze2023hcu} for discussion.  

Spin-glass ordering is induced by a transverse pump at $\lambda=780$~nm.  It drives the atoms with Rabi frequency $\Omega$ and atomic detuning $\Delta_A = -2\pi{\cdot}97.2$~GHz. This pump is retroreflected to form a standing wave ${\propto}\cos(k_rx)$, where $k_r = 2\pi/\lambda$ and $k\approx k_r$ above. The pump frequency is red-detuned from the $\eta=0$ family by $\Delta_C = -2\pi{\cdot} 20$~MHz~\footnote{The other $\eta$ families are detuned by ${>}1$~GHz and negligibly contribute.}. A superradiant phase transition occurs at a critical pump power ${\propto}\Omega^2_c$. We exponentially ramp the pump from zero to $\Omega^2 = 4\Omega^2_c$ over a time $t_R$, as shown in Fig.~\ref{fig:Fig1}b. This is followed by a quench to ${\sim}5\Omega^2_c$ for increasing imaging signal over a time  $t_q = 300$~$\mu$s. 

The superradiant transition is describable by a multimode version of the Hepp-Lieb-Dicke model~\cite{Kirton2018itt,Kroeze2023rsb,Supp}.   Above threshold, each atomic ensemble forms either one of two checkerboard-like density waves, defining the  ``up'' and ``down" effective spins.  Concomitantly, the up (down) spins superradiantly scatter light into the cavity with phase $\pi$ (0) with respect to the pump. The emitted cavity field is imaged with a camera for holographic spin readout, and we plot the up (down) spins blue (orange). 

The atomic density waves are captured by a two-mode description that leads to a transverse-field Ising model after adiabatic elimination of the cavity modes~\cite{Supp},
\begin{equation}\label{eq:Ham}
    \hat{H} = 2E_r \sum_{i=1}^n \hat{S}_i^z - \frac{\hbar g_0^2\Omega^2}{\Delta_A^2|\Delta_C|}\sum_{ij=1}^n J_{ij}\hat{S}_i^x \hat{S}_j^x 
\end{equation}
where $E_r=\hbar^2k_r^2/2m$ is the recoil energy, $\hat{S}_i^{x/y/z}$ are collective spin operators of total spin $S=N_A/2$ for each ensemble, and $m$ is the atomic mass.  Together with Lindbladian dissipation~\cite{Supp}, this realizes a nonequilibrium variant of the SK model where the ordering transition is driven by a competing transverse field rather than thermal fluctuations. 

The connectivity matrix of the interaction is given by the cavity Green's function, 
$J_{ij}\propto\delta(\mbf{r}_i-\mbf{r}_j) + G^{\eta}_\mathrm{non}(\mbf{r}_i,\mbf{r}_j)$;
see Ref.~\cite{Supp} for full version including  effects of imperfect degeneracy and finite atomic ensemble width. The delta-function term encourages the atoms within each ensemble to order with the same density wave phase and arises from the constructive superposition of many Hermite-Gaussian modes. The nonlocal term arises from three defocused components of the cavity field at its midplane:
\begin{equation}\label{eq:Greens}
    G^{\eta=0}_\mathrm{non}(\mbf{r}_i,\mbf{r}_j)= \sum_{\nu=1}^3\frac{1}{\pi s_\nu}\sin\Big[ \frac{2\nu\pi}{7} + \frac{c_\nu\big(\mbf{r}_i^2 + \mbf{r}_j^2\big)-2\mbf{r}_i\cdot\mbf{r}_j}{s_\nu w_0^2}\Big], 
\end{equation}
where $c_\nu=\cos(2\nu\pi/7)$, $s_\nu=\sin(2\nu\pi/7)$, and the waist of the $\Xi_{00}$ mode is $w_0 = 35$~$\mu$m. This interaction can change sign depending on the positions of the atomic ensembles. 
Indeed, trapping them roughly $w_0$ apart randomizes the $J_{ij}$ signs. This form of quenched disorder induces spin frustration~\cite{Marsh2021eam,Supp} in a manner akin to the SK model.

\begin{figure*}[t!]
    \centering
    \includegraphics[width=\linewidth]{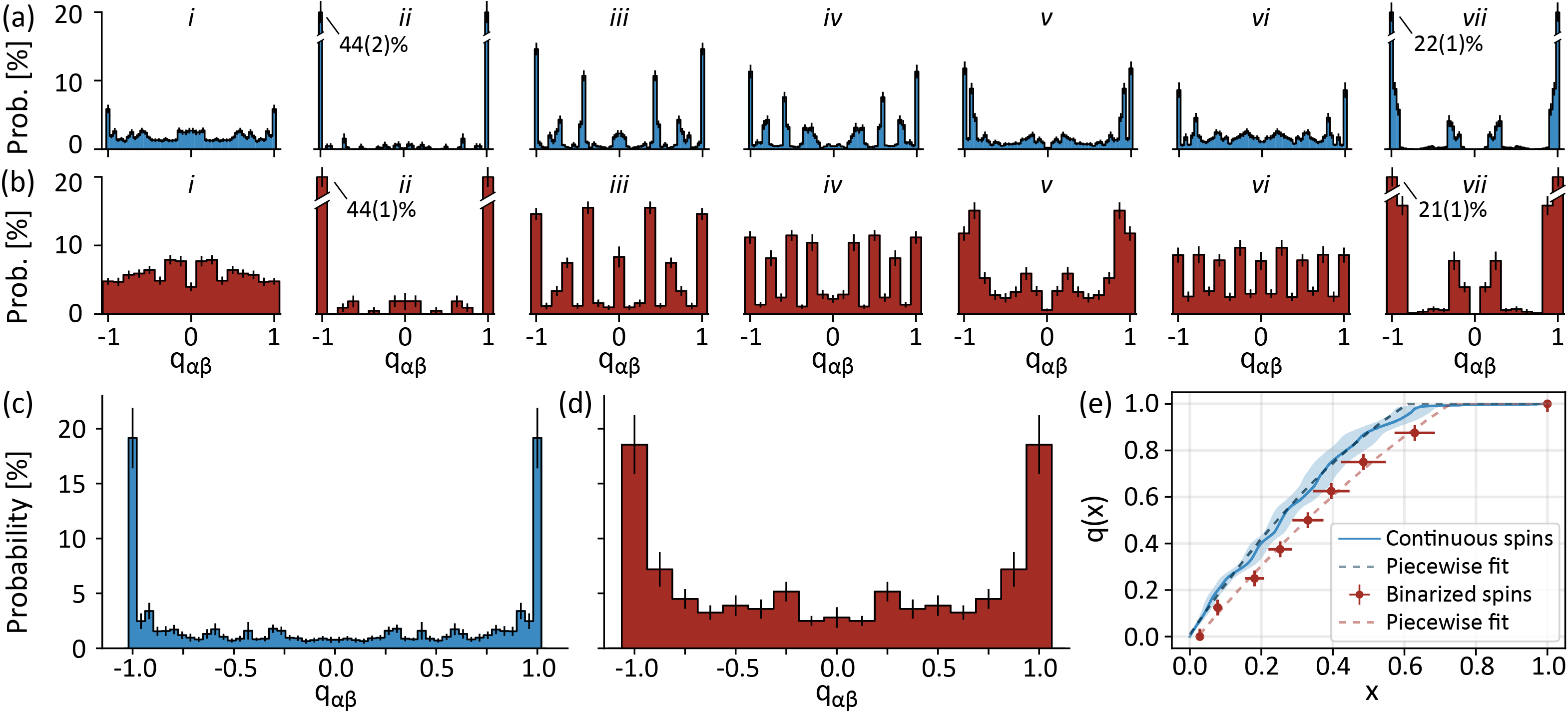}
    \caption{(a) Gallery of overlap distributions for seven disorder realizations of the coupling matrix, $J_1$--$J_7$, for $n= 16$. (b) Overlap distributions for these same disorder realizations after spin binarization. (c) The Parisi distribution of all 14 disorder realizations. (d)  Parisi distributions after spin binarization. (e) Parisi functions for the continuously valued spins (blue) and spins after binarization (red). Dashed lines show least-squares fits to linear-constant piecewise functional forms. }
    \label{fig:Fig3}
\end{figure*}

To determine whether the 4/7 cavity QED system actually yields a spin glass, we first perform experiments focused on a single disorder instance ``$J_1$" that couples $n=16$ spins; see Ref.~\cite{Supp} for atomic ensemble locations. The all-to-all connectivity diagram for $J_1$ is plotted in Fig.~\ref{fig:Fig2}(a). Each experiment begins with the atomic ensembles polarized along $\ex{\hat{S}^z_i}  = -S$.  Pumping the system beyond the ordering transition results in the organization of the spins into a configuration $\mbf{s}^\alpha = ( \langle \hat{S_1^x} \rangle,\cdots, \langle \hat{S}_n^x \rangle ) /\mathcal{N}$, where the normalization $\mathcal{N}$ is chosen such that $\mbf{s}^\alpha$ has unit norm, removing the effect of global atom number fluctuations.  Repeating this many times---i.e., creating new ultracold gases at the same locations to realize the same $J_1$, and pumping in the same way through the transition---yields a set of ``replica'' spin configurations of the glass, each indexed by $\alpha$~\cite{Marsh2024ear,Kroeze2023rsb}. 

Processed cavity emission images from three replicas are shown in Fig.~\ref{fig:Fig2}b-d; Ref.~\cite{Supp} presents image analysis methods. Each bright spot is the local electric field emitted by one of the atomic ensembles.
The sign and amplitude of the local fields are fit to yield a measurement of each spin amplitude $\langle \hat{S_i^x} \rangle$, where the camera integration time $t_q$ implements the average.  A histogram of these is presented in Fig.~\ref{fig:Fig2}e, showing a bimodal distribution. This allows us to delineate spins with $\langle \hat{S_i^x} \rangle > 0$ (${<}0$) as being up (down).  Rarely, the field emitted by an atomic ensemble splits into two components of opposite phase.  Panel (b) shows an example of one such split ensemble in the third row and column; see Ref.~\cite{Supp} for discussion of this frustration-reducing effect and the fit routine.  Such cases result in the group of $\langle \hat{S_i^x} \rangle$ near zero in panel (e).

The ramp time $t_R$ controls the variety of spin configurations observed. Figure~\ref{fig:Fig2}f shows the base-2 Shannon entropy~\cite{Supp} of the spin configurations resulting from ensembles of 200 replicas taken at different $t_R$. For simplicity, only the sign of the spins are considered when computing the entropy. We find that short ramp times yield a large number of distinct spin states, and therefore high entropy, while longer ramps yield fewer, but more frequently observed states. This suggests that in this finite-size system, slow ramps permit spin organization into only a few states. Indeed, a similar measurement in the $n=8$ vector spin glass showed that a single state can be reached with slow ramps~\cite{Kroeze2023rsb}.

\begin{figure}[t!]
    \centering
    \includegraphics[width=\columnwidth]{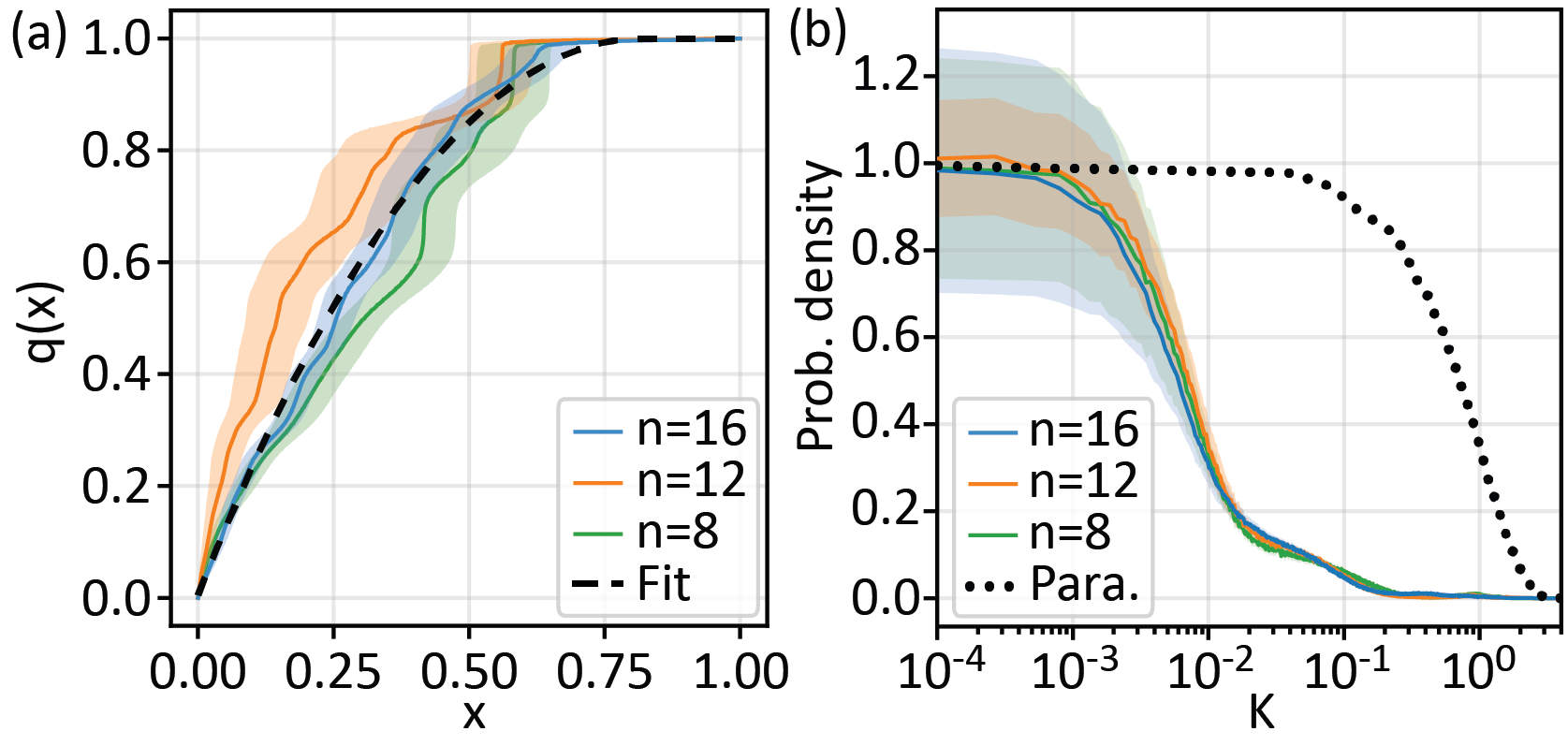}
    \caption{(a) Measured Parisi function $q(x)$ for system sizes $n=8$, 12, and 16 versus the cumulative overlap probability $x$. (b) Ultrametricity $K$-correlator distributions, normalized to the peak probability density of each system size. A comparison to that of the paramagnetic phase is provided.}
    \label{fig:Fig4}
\end{figure}

A central manifestation of RSB is the formation of correlations among replica spin configurations. Parisi showed that the replica overlap $q_{\alpha\beta}=\mbf{s}^\alpha{\cdot}\mbf{s}^\beta$ correlator should exhibit a fractal, block-diagonal matrix form~\cite{Parisi1980top}, an idealized example of which is shown in Fig.~\ref{fig:Fig1}e. Moreover, the overlap distances between replicas, $d_{\alpha\beta}=1-|q_{\alpha\beta}|$, form an ultrametric space asymptotically in system size. Ultrametric spaces satisfy the strong triangle inequality $d_{\alpha \beta} \leq \max\{{d_{\alpha \gamma}, d_{\beta \gamma}}\}$ in the overlap distance. This may be visualized as a family-tree-like dendrogram, as shown in Fig.~\ref{fig:Fig1}d.  Microscopic spin readout allows us to directly measure these quantities: Figs.~\ref{fig:Fig2}g show the hierarchical clustering of 200 replicas for ramp times varying from 0.1~ms to 20~ms, while Figs.~\ref{fig:Fig2}h show the associated full overlap matrices $q_{\alpha\beta}$. We do indeed observe the formation of nested clusters of correlated replicas with cluster sizes controlled by the ramp time. We will return to a more quantitative analysis of this nascent ultrametric structure in Fig.~\ref{fig:Fig4}. 

The overlap distributions plotted in Figs.~\ref{fig:Fig2}i are histograms of $q_{\alpha\neq\beta}$~\footnote{The overlap distribution is discretized into 50 bins and symmetrized over the global $\mathbb{Z}_2$ symmetry $\mbf{s}^\alpha{\to}-\mbf{s}^\alpha$.}. A primary feature is the emergence of ``goalpost" peaks near $q_{\alpha\beta}=\pm 1$. These peaks are a consequence of low spin-ensemble entropy, which causes replicas to cluster around common spin configurations, and indicates an ordered phase.  The formation of additional internal structure and peaks between the goalposts is direct evidence of RSB~\cite{Stein2013sga}; the continuous distribution we observe is due to finite network size. Note that magnetization is also zero~\cite{Supp}.  To the best of our knowledge, this is the first microscopic observation of spin states of a deeply ordered Ising spin glass.  Glassy overlap structure persists throughout a range of ramp times, though the system approaches a paramagnetic state at short $t_R$. We avoid this by employing $t_R=5$~ms below. Note that 400 replicas at $n=25$ have been measured and exhibit qualitatively similar overlap distributions~\cite{Supp}. 

We now create different experimental disorder realizations of $n=16$ spin glasses. The measured overlap distributions for seven disorder realizations, $J_1$ through $J_7$, are shown in Figs.~\ref{fig:Fig3}a; Ref.~\cite{Supp} presents seven others. The absence of self-averaging in spin glasses results in distinct distributions between the goalposts~\cite{Stein2013sga}. In accordance with this expectation, we observe that the interior peak structure varies between the realizations.  An order parameter that is independent of sample-to-sample differences is the Parisi distribution, the disorder average of the overlap distribution. Figure~\ref{fig:Fig3}c presents an average of 14 disorder realizations, each derived from 200 replicas. Despite the intrinsic nonequilibrium nature of the driven-dissipative, experimental spin glass, the measured Parisi distribution is qualitatively consistent with the equilibrium expectation from the SK model~\cite{Stein2013sga,Parisi1980aso}: i.e., a filled interior distribution between $q_\text{EA}\equiv q_{\alpha\alpha}$ goalposts. This indicates a well-ordered experimental spin glass phase.  

Both natural~\cite{Binder1986sge} and theoretical~\cite{Sherrington1975smo} spin glasses can arise from binary spin degrees of freedom. While the spins of our system are continuous in amplitude, the bimodal amplitude distribution in Fig.~\ref{fig:Fig2}e suggests their binarization may retain essential correlations while allowing for a simplified model description. 
Figures~\ref{fig:Fig3}b show the overlap distributions after spin binarization, where only $n+1=17$ overlap values are now possible. Qualitative features persist, with peaks appearing in similar locations and with similar amplitude~\footnote{Note that the large bins cause the goalposts to appear lower in amplitude even though they are of similar total probability.}.  This admits the possibility of using a binarized description of the physical system, e.g., in associative memory applications~\cite{Marsh2025eam}.

The Parisi function $q(x)$ is a quantity related to the Parisi distribution and plays an essential role in the theory of RSB regarding the structure of the disorder-averaged replica overlap matrix~\cite{Parisi1980top}. We extract $q(x)$ from our measurements as follows. The cumulative probability density of the disorder-averaged $|q_{\alpha\beta}|$ distribution is $x(q)$.  Taking the function inverse, one obtains $q(x)$, which is the overlap value $q$ for which a fraction $x$ of the Parisi distribution is contained between $\pm q$. A non-constant functional form of $q(x)$ indicates RSB. $q(x)$ takes a piecewise linear-constant form in the SK model near the critical temperature while developing a nonlinear-constant form deeper into the spin glass phase~\cite{Parisi1980top,Parisi1980aso}. 
While ``$k$-step'' RSB corresponds to $q(x)$ functions that increase in $k$ discrete steps before plateauing at $q(x)=q_\text{EA}$, the SK model exhibits ``full" (i.e., infinite-step) RSB in which $q(x)$ smoothly increases to $q_\text{EA}$. Despite the presence of continuously varying spin amplitudes and nonequilibrium dynamics, the measured Parisi functions in Fig.~\ref{fig:Fig3}e fit well to a piecewise quadratic-constant functional form, similar to the thermal equilibrium SK model. This is true both with and without binarization of the spins, and both $q(x)$ are similar within error. We extract a $q_\text{EA}=0.99$ (0.98) for the continuously varying (binarized) spins using least-squares fits. These $q(x)$ constitute direct evidence of full RSB in experimental driven-dissipative spin glasses. 

Figure~\ref{fig:Fig4}a shows $q(x)$ for system sizes $n=8$, 12, and 16; see Ref.~\cite{Supp} for parameter sets. We find these $q(x)$ to be qualitatively similar, both regarding slope and their near-unity saturation of $q_\text{EA}$. All three are consistent (within the error each) of a single piecewise quadratic-constant form with $q_\text{EA}=0.999(1)$. We further investigate the degree to which ultrametricity varies versus $n$. The $K$-correlator~\cite{Katzgraber2009uac,Supp} quantifies the degree to which the strong triangle inequality is satisfied by generating a distribution of $K$ over triples of replicas. The $K$-distribution approaches a delta function at $K=0$ for a perfectly ultrametric space but broadens due to violations of this inequality from, e.g., finite-size effects. Figure~\ref{fig:Fig4}b shows that the $K$ distribution assumes a common form across the measured system sizes with full width at half maximum (FWHM) of $0.007(1)$ and mean $\ex{K}=0.25(1)$. Comparison to the paramagnetic limit~\cite{Supp} with FWHM of $0.72(4)$ and $\ex{K}=0.66(1)$ shows a significant degree of ultrametricity at all sizes. While the $K$-distribution does not seem to narrow with increasing $n$, we note that this observation is consistent with the very slow emergence of ultrametric correlations exhibited by the SK model, as studied using finite-size scaling simulations~\cite{Katzgraber2009uac}.

In summary, we have characterized an Ising spin glass in a quantum-optical setting by measuring  replicas at the individual spin level. The 4/7-cavity QED spin glass enables the creation of associative memory~\cite{Marsh2025eam}.  Future work will explore glassy dynamics as well as extensions into the spin-1/2 limit of quantum spin glasses using Rydberg-blockaded ensembles~\cite{Marsh2024ear}; this will allow comparisons to quantum SK theories~\cite{Tikhanovskaya2024edo,Kiss2024crs}.

We thank Ronen Kroeze, Yudan Guo, and Sarang Gopalakrishnan for valuable discussions. We are grateful for funding support from the Army Research Office (Grant \#W911NF2210261) and the Q-NEXT DOE National Quantum Information Science Research Center. 
J.K.~acknowledges support from EPSRC (Grant No.~EP/Z533713/1).
B.M.~acknowledges funding from the Stanford QFARM Initiative.
H.H.~acknowledges support from the Stanford Shoucheng Zhang Graduate Fellowship.

\clearpage
\onecolumngrid

 \let\oldaddcontentsline\addcontentsline
        \renewcommand{\addcontentsline}[3]{}

\section*{Supplementary Information}

        \let\addcontentsline\oldaddcontentsline

\tableofcontents

\renewcommand{\theequation}{S\arabic{equation}}
\renewcommand{\thefigure}{S\arabic{figure}}
\setcounter{figure}{0}
\setcounter{equation}{0} 
\setcounter{figure}{0}
\setcounter{equation}{0} 

\section{Experimental methods}

\subsection{Ultracold atom production}
\label{ColdAtomProd}

The production of an ultracold gas of $^{87}$Rb initially follows the procedures described in previous work~\cite{Kollar2015aac,Kroeze2023rsb}. In summary, atoms are loaded into a magneto-optical trap over $6.1$~s followed by polarization gradient cooling and radio frequency (RF) evaporation, cooling $1.3(2)\times 10^8$ atoms in the $\ket{F,m_F}=\ket{1,-1}$ hyperfine state to approximately $40\,\mu$K~\cite{Kollar2015aac}. A 1064-nm optical dipole trap (ODT) then loads $3.6(4) \times 10^7$ atoms and transports them to the approximate center of the multimode cavity, resulting in $2.3(2)\times 10^7$ atoms. The atomic cloud is then handed off to an orthogonal pair of 1064-nm crossed ODTs (XODTs) that lie in the midplane of the vertically oriented cavity. Each XODT beam has a waist of 21~$\mu$m. The transverse positions of the XODTs are controlled using acousto-optic modulators (AOMs) driven at an RF frequency centered at 80~MHz. The AOMs are first driven by a pair of voltage controlled oscillators (VCOs) and later by an arbitrary waveform generator (AWG) for splitting the gas into multiple sites~\cite{Kroeze2023msp}.

For this work, a new splitting procedure has been designed to generate arrays of up to $5\times5$ sites. Once the atoms are transported to the cavity center by the ODT, the XODT1 and XODT2 beams are ramped on over $100$~ms and dithered by driving the VCOs with 6~kHz and 10~kHz triangle waveforms, resp. This produces a painted potential corresponding to a two-dimensional flat-bottom trap of size $64\,\mu$m$\times64\,\mu$m. Next, the ODT power is linearly ramped off over 2.45 seconds. During this time, the XODT powers are also ramped using an exponential decay to optically evaporate the atoms. The result is $1.55(9)\times 10^6$ atoms cooled to near the critical temperature $T_c$ for Bose-Einstein condensation.

Cold atoms in the flat-bottom trap are then split into an array of $n_x\times n_y$ sites using time-dependent RF waveforms generated by the AWG with two independent output channels (one for each XODT) operated with a 500 MHz sampling rate.  The AWG waveforms are calculated for each new experimental sequence to allow for shot-to-shot position stabilization as described in Sec.~\ref{sec:positions}.  We found that this sampling rate provides sufficient frequency resolution while allowing the numerical computation of the waveforms to occur within the experimental cycle time of $\approx$ 23~s~\cite{Marsh2024qnn}.

\begin{figure}
    \centering
    \includegraphics[width=\textwidth]{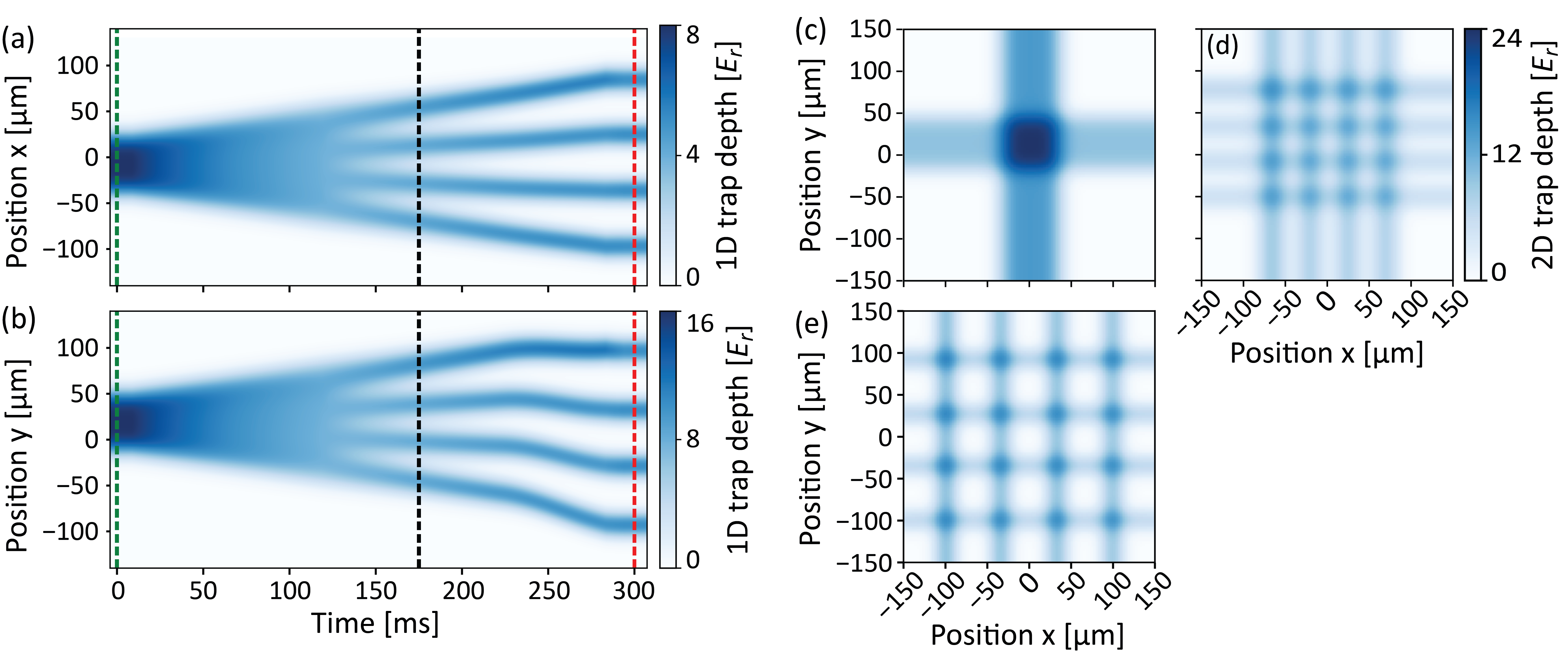}
    \caption{ Visualization of the time-dependent trap depth for the preparation of an array of $n_x=4$ by $n_y=4$ sites. Trap depth is shown in units of the recoil energy $E_r$ for $^{87}$Rb at 1064~nm. 
    (a) The normalized optical potential generated in the $x$-direction and (b) in the $y$-direction. The 2D potential is the sum of the two trap depths in each direction. (c) The initial 2D flat-bottom trap potential at $t=0$ ms (green dashed lines) corresponds to the end of the optical evaporation sequence. (d) By $t=175$ ms (black dashed lines), the 2D flat-bottom trap has grown in size and started to develop separated wells, thereby splitting the atomic cloud into localized sites. (e) The final trap shape at $t=300$ ms (red dashed lines), resulting in cold atom clouds arranged in a rectilinear array at the intersections of the XODT beams.}
    \label{SuppFig:XODTTrajectory}
\end{figure}

The splitting sequence begins by switching the AOM RF source from the VCOs to the AWG with an RF switch. The AWG-generated potential then evolves to split the flat-bottom trap into individual trapping sites as follows: 1) The potential initially matches the $64\,\mu$m$\times64\,\mu$m flat-bottom trap generated by the VCOs during the handoff of the RF source. 2) The flat-bottom trap is linearly expanded to $127\,\mu$m$\times127\,\mu$m over ${\sim}$125~ms to allow room for the sites to form in each direction. 3) This enlarged trap is partitioned into $n_y$ sections in the XODT1 direction and $n_x$ sections in the XODT2 direction. Each section of the trap is then translated away from the others over ${\sim}125$~ms. The precise size of each partition can be adjusted to balance the atomic populations in each site. Simultaneously, the amplitude of the dithering triangle wave used to generate the flat-bottom trap is linearly ramped off. By the end of the separation process, each channel of the AWG is producing pure RF tones, each corresponding to a trapping site, time-multiplexed at 10~kHz. The trap sites are then translated over $\sim50$~ms to the desired final positions in the cavity midplane. During this process, the trap depth experienced by each cloud of atoms is reduced so that the final clouds approach the critical Bose-condensation temperature, as measured via a time-of-flight (TOF) technique.

The total time-dependent trap depth for each direction is illustrated in Fig.~\ref{SuppFig:XODTTrajectory}, accounting for the waist of the XODT beams. After the aforementioned splitting procedure, the atoms remain trapped at the intersections of the beams for the remainder of the experiment.  These intersections are illustrated in Fig.~\ref{SuppFig:XODTTrajectory}e. The initial flat-bottom trap is partitioned with nonuniform trap widths to adjust atomic populations between sites. This is done to balance the fitted spin amplitudes rather than the raw atom numbers. We are able to achieve 90\% uniformity in average spin amplitude per row and column by tuning the trap widths during this step. The nonuniform trap widths result in nonuniform trap depths across sites at the end of the splitting procedure. A rebalancing step is then performed over 10~ms to approximately match the trap depths. The total power of all traps is then ramped up over 20~ms to more tightly confine the atoms. The XODT1 laser power is increased by a factor of 4 while the XODT2 laser power is increased by a factor of 6.84. This produces atomic clouds that are nearly isotropic in shape as measured in cavity emission images; see Supp. Sec.~\ref{sec:ImageAnalysis}. Thus we obtain an array of cold atomic clouds in a rectilinear grid of programmable size and spacing. 

Full characterization of the atom number, temperature, and trap depth for each site, for each disorder realization, is prohibitively time-consuming and only indirectly related to the spin glass physics of interest in this work. We have nonetheless measured these quantities for disorder realization $J_1$. We measure an atom number of $6(1){\times}10^4$, where parentheses indicate standard deviation across sites. The mean temperature is $0.64(5)$~$\mu$K, the mean BEC fraction is $5(2)\%$, and the mean trap frequencies are $[\omega_x,\omega_y,\omega_z]=2\pi\cdot[326(5),472(16),332(9)]$~Hz.  Additional atom number and temperature measurements are performed for $J_3$ (system size $n=16$). We find an average atom number of $5(1){\times}10^4$, an average temperature of $0.62(5)$~$\mu$K, and an average BEC fraction of $8(5)\%$.  

Atom numbers are measured by absorption imaging after a fixed TOF. Site-resolved measurements of the trap frequencies, atom number, and Bose-Einstein condensate (BEC) fraction requires isolating a single site. To do this, the atoms are first split into the particular configuration realizing the disorder configuration to be measured, and then the trap beams of the unwanted sites are rapidly redirected to a far-away region. This pulls the optical potentials out from under the atoms in the unwanted sites, allowing them to fall away under gravity. Waiting 12~ms in this configuration is sufficient to prevent the atoms from being imaged at the position of the single desired site. The trap then reverts to the original $n_x$ by $n_y$ array, resetting all the prior trapping conditions. The single site of interest is then measured individually in TOF imaging. This site-resolved atom number measurement is used to obtain the relative number of atoms per site. We use a camera with higher quantum efficiency to measure the total number of atoms across all sites through non-site-resolved imaging. Together these measurements yield site-resolved absolute atom numbers. 
BEC fractions are measured by fitting TOF absorption images to the sum of a Gaussian and a Thomas-Fermi distribution. Trap frequencies are measured through the imaging of atomic oscillations in the trap. This measurement is carried out for all sites of $J1$; we did not repeat for $J3$, but they are expected to be similar because the final power per trap is balanced. The BEC fractions combined with the measured trap frequencies suffice to approximate the temperature of the gas at each site. Large thermal fractions shift the superradiant transition point~\cite{Piazza2013bcv}, but do not otherwise affect the physics of the spin glass presented here. In the absence of transverse pumping, the $1/e$ lifetime of the atoms trapped in the rectilinear array decay is 4.4~s, mainly limited by three-body loss and vacuum background collisions~\cite{Kollar2015aac}. 

\subsection{Selection and stabilization of spin locations}
\label{sec:positions}

Disorder realizations of the coupling matrix $J$ are created by placing the atomic ensembles in different locations within the cavity midplane. For each system size $n$ there are an associated number of rows $n_y$ and columns $n_x$ corresponding to the traps described in Sec.~\ref{ColdAtomProd}. Spins are trapped at the intersections of these rows and columns at locations $(x_i,y_j)$, where $\{x_1,\cdots,x_{n_x}\}$ are the column locations and $\{y_1,\cdots,y_{n_y}\}$ are the row locations referenced to the cavity center. The column locations are parameterized by 
\begin{equation}
    x_i = x_c + \left(i-\frac{n_x-1}{2} \right)d_x + \delta_{i,x},
\end{equation}
where $x_c$ is a center position parameter, $d_x$ is a nominal spacing parameter between columns, and each $\delta_{i,x}$ is a shift applied to each column. The rows are similarly parameterized by 
\begin{equation}
    y_j = y_c + \left(j-\frac{n_y-1}{2} \right)d_y + \delta_{j,y},
\end{equation}
where the parameters $y_c$, $d_y$, and $\delta_{j,y}$ play analogous roles for the row locations.

\begin{table}[t]
  \centering
  \begin{tblr}{
      colspec={cccccccccc},
      row{even}={bg=gray!10},
    }
     Group & $n$ & $n_x$  & $n_y$  & $d_x$ ($\mu$m) & $d_y$ ($\mu$m)& $w_\mathrm{c,x}$ ($\mu$m) & $w_\mathrm{c,y}$ ($\mu$m) & $w_x$ ($\mu$m) & $w_y$ ($\mu$m)\\
    A & 16 & 4 & 4 & 62    & 62   & 14   & 14 & 6  & 6 \\
    B & 12 & 3 & 4 & 85    & 62   & 18   & 14 & 18 & 6 \\
    C & 8  & 2 & 4 & 130   & 62   & 26 & 14 & 50 & 6 \\
    D & 8  & 4 & 2 & 62    & 124  & 14   & 14 & 6  & 6 \\
  \end{tblr}
  \caption{Parameters used to randomly generate the spin positions for each disorder realization of $J$ matrix. See text for parameter descriptions.}
    \label{SuppTab:PosParams}
\end{table}

We use the following method to find a set of row and column locations for each disorder instance.  All parameters are given in Table~\ref{SuppTab:PosParams} for each of the four groups of settings employed.  Group A contains the parameters for all 14 disorder realizations of $J$ matrix studied for $n=16$. Group B contains all 10 disorder realizations for $n=12$. Groups C and D both correspond to $n=8$ but with opposite $n_x$ and $n_y$ to diversify the set of spin positions. The center coordinates $(x_c,y_c)$ are sampled from uniform random distributions for each disorder instance in the ranges $x_i\in[-w_\mathrm{c,x}/2,w_\mathrm{c,x}/2]$ and $y_j\in[-w_\mathrm{c,y}/2,w_\mathrm{c}/2]$. The individual row and column shifts are also sampled from uniform random distributions: $\delta_{i,x}\in[-w_{x}/2,w_x/2]$ and $\delta_{j,y}\in[-w_y/2,w_y/2]$. 

The parameters are chosen to maximize the diversity of spin positions while conforming to experimental constraints. The maximum deviation from the cavity center is kept less than 150~$\mu$m to avoid effects arising from cavity aberration. The minimum separation between any two atomic ensembles is kept greater than $40\mu$m to ensure the trap potentials from separate rows and columns are well separated. These parameters also permit repeatable performance during the splitting procedure described in Sec.~\ref{ColdAtomProd}. The average distance from the cavity center is held at $93~\mu\mathrm{m}\approx 2.7 w_0$ across system sizes; recall that the Gaussian mode waist is $w_0 = 35$~$\mu$m. This degree of position spread leads to highly frustrated $J$ matrices similar to that of the SK model; see Ref.~\cite{Marsh2021eam} and Sec.~\ref{sec:eigs} for the relation between position spread and spin frustration. The positions for $J_1$, the disorder instance used in the main text, are:   $x$-positions, $x_i\in\{-97.15, -36.3, 25.2, 85.1\}$~$\mu$m; and $y$-positions, $y_j\in\{-93.4, -28.9, 32.3, 97.3\}$~$\mu$m.

Shot-to-shot position feedback is used to stabilize the spin positions against a slow drift over time due to, e.g., thermal fluctuations in optical components. A five-shot running average of the center-of-mass (COM) spin position is extracted from the fit routine described in Sec.~\ref{sec:ImageAnalysis}. Deviation of the five-shot-averaged position from the intended locations $\bar{x}=\sum_i^{n_x} x_i/n_x$ and $\bar{y}=\sum_j^{n_y} y_j/n_y$ serves as a feedback signal to the tones generated by the AWG. A correction to the AWG proportional to the COM deviations is applied between every shot to stabilize the atomic positions. Experimental shots with a fitted COM that deviates by more than $1~\mu$m from $(\bar{x},\bar{y})$ are excluded. This results in average rejection ratios of $7.5\%$, $6\%$, and $11\%$ for $n=16$, $12$, and $8$, resp. After rejecting outliers, the remaining shots are position-stabilized to within ${<}0.5$~$\mu$m average COM deviation.

\subsection{Multimode optical cavity} 

This work features an optical resonator of a different length and mode structure compared to our previous work~\cite{Kroeze2023rsb,Kollar2015aac}. The mirrors remain the same with an $R=1$~cm radius of curvature but their separation has been increased from the 1-cm distance necessary for a confocal configuration, to $L\approx 1.22$~cm.  This $L/R$ ratio nearly coincides with the multimode degeneracy point $L/R=2\sin^2(M\pi/2N)$~\cite{Guo2019eab} for the irreducible fraction $M/N=4/7$.  ($M/N=1/2$ for the confocal cavity.) To the best of our knowledge, this is the first cavity QED experiment employing an $M/N=4/7$ resonator. As discussed in the main text, a primary advantage of generalized $M/N$ resonators is the ability to tune cavity-mediated spin-spin interactions between a vector form, such as $J_{ij}(X_iX_j-Y_iY_j)$ for confocal cavities, to the Ising $J_{ij}X_iX_j$ form realized by the $4/7$ cavity. 

\begin{figure}[t]
    \centering
    \includegraphics[width=\linewidth]{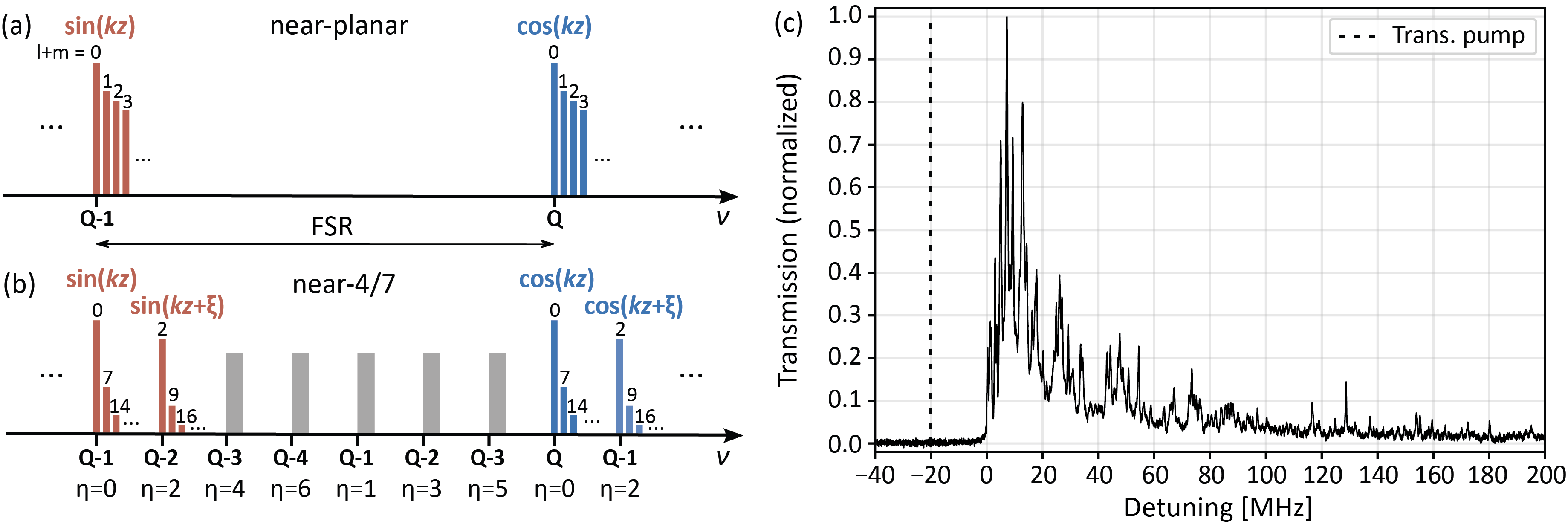}
    \caption{(a) Diagram of the mode spectrum of a near-planar cavity compared to that of (b) a near-$4/7$ cavity. While the near-planar has only mode family per FSR which contains all HG modes, the near-$4/7$ cavity has $N=7$ families with constant $\eta=l+m\mod N$ per FSR. Each family is labeled by its longitudinal mode number $Q$ and $\eta$. Color indicates longitudinal quadrature up to a phase shift $\xi=\eta M\pi/(2N)$. (c) Experimental transmission spectrum of the $4/7$ cavity. The vertical dashed line shows the frequency detuning $\Delta_C=-2\pi\cdot 20$~MHz of the transverse pump.}
    \label{fig:cavSpec}
\end{figure}

We now present essential information about $M/N$ resonators while leaving analytic derivations of the cavity-mediated interactions to Sec.~\ref{sec:theory}. An $M/N$ multimode cavity supports degenerate mode families composed of different subsets of Hermite-Gaussian (HG) modes $\Xi_\mu$. Each HG mode is indexed by its transverse mode indices $\mu=(l,m)$ and its longitudinal mode number $Q$. In brief, the number $N$ in the $M/N$ designation describes which HG modes belong to which family. The number $M$ describes the tradeoff between longitudinal and transverse mode numbers within a family~\cite{Guo2019eab}. More specifically, an $M/N$ mode family contains all those HG modes with $(Q,l+m)=(Q_0-iM,\eta+iN)$ for integers $i\geq0$, where $Q_0$ is the longitudinal mode number of the lowest-order transverse mode in the family and $\eta=l+m\mod N$ is constant within the family. The numbers $Q_0$ and $0\leq\eta<N$  uniquely determine a mode family. The resonance frequency of a perfectly degenerate $(Q_0,\eta)$ family is
\begin{equation}
    f_{Q_0,\eta}=\frac{c}{2L}\left(Q_0+\frac{M}{N}(1+\eta) \right),
\end{equation}
where $c$ is the speed of light. From the above, we see that the resonance frequencies of $M/N$ families are uniformly spaced in frequency by $c/(2NL)$, or $1/N$ of the free spectral range (FSR). The families are not sequential in $\eta$, in general; Fig.~\ref{fig:cavSpec}a diagrams the mode spectrum of the $4/7$ cavity and the location of each family. 

The longitudinal behavior of modes within a family depends on $M$. The HG modes have a longitudinal form that varies cyclically with $Q$ as $\sin(kz+\xi),\cos(kz+\xi),-\sin(kz+\xi),-\cos(kz+\xi)$ near the cavity midplane, where $k$ is the mode wavevector and $\xi=\eta M\pi/(2N)$. Since the modes within a family have $Q=Q_0-iM$ for integers $i$, the parity of $M$ plays an important role: Families of odd-$M$ cavities contain degenerate modes with both $\sin(kz+\xi)$ and $\cos(kz+\xi)$ form, while the degenerate modes within families of even-$M$ cavities have fixed longitudinal forms of either $\sin(kz+\xi)$ or $\cos(kz+\xi)$. This distinction is utilized here to realize Ising interactions; see Sec.~\ref{sec:theory} for details.

In our experiments, we couple to an $\eta=0$ family in a 4/7 resonator. It contains the Gaussian fundamental mode $\Xi_{0,0}$ and all other HG modes $\Xi_{l,m}$ for which $l+m=0\mod 7$. Its experimental transmission spectrum is shown in Fig.~\ref{fig:cavSpec}b versus the frequency of a probe laser focused to a spot with a Gaussian waist of 8~$\mu$m located in the center of the cavity. Imperfect mode degeneracy results in a mode bandwidth on the order 20-40~MHz. The spectrum depends on the shape of the probe beam, since different beam shapes couple to different spatial patterns of the intracavity field.  In the present experiments, we choose a cavity length slightly longer than the nominal length determined by the degeneracy condition $L=2R\sin^2(M\pi/2N)$.  This is done so that the lowest-frequency mode is the $\Xi_{00}$ fundamental mode, with all higher-order modes farther red detuned from the pump frequency $\omega_P$. This allows the cavity to be well described by the Green's function in Sec.~\ref{sec:Greens}. We choose to define the cavity detuning as $\Delta_C=\omega_P-\omega_{0,0}$ with respect to the $\Xi_{00}$ mode. This is the first peak in the transmission spectrum, as was also the case in our previous work~\cite{Vaidya2018tpa,Kroeze2023rsb}. 

We now discuss the cavity QED parameters. The FSR is $2\pi\cdot 12.30010(6)$~GHz. This leads to an estimated finesse of $44,040(40)$ from the $\Xi_{00}$ field decay rate of $\kappa=2\pi\cdot 140(1)$~kHz. 
We have previously found that $\kappa$ varies little for other HG modes~\cite{Kollar2015aac}. The single-mode, single-atom coupling strength $g_0$ was previously measured to be $2\pi\cdot1.47$~MHz for this optical cavity in the confocal configuration~\cite{Kroeze2023hcu,Vaidya2018tpa}. The coupling strength scales as $1/\sqrt{V}$, where $V=\pi w_0^2L/4$. The mode waist $w_0$ decreases from 35.2~$\mu$m in the confocal configuration to 34.8~$\mu$m in the $4/7$ configuration. This decreases the estimated single-atom, single-mode coupling strength of the $4/7$ cavity to $g_0=2\pi\cdot 1.35$~MHz. The single-mode cooperativity of the $4/7$ cavity is thus $C=g_0^2/(\kappa \gamma_\perp)=4.28$, where the spontaneous emission rate is $\gamma_\perp=\Gamma/2$ and $\Gamma=2\pi\cdot 6.0659$~MHz for the $^{87}$Rb D$_2$ transition. The multimode nature of the cavity boosts the effective cooperativity in the dispersive limit of cavity driving, yielding an enhancement of up to a factor of $21$ for the confocal configuration~\cite{Kroeze2023hcu}. However, the number of degenerate HG modes of the same longitudinal quadrature decreases by a factor of $4/7$ going from the confocal cavity to the $4/7$ cavity. This leads to an estimated multimode enhancement of 6.85 and an estimated dispersive-limit multimode cooperativity of $C_\mathrm{mm}=29$ for the $4/7$ cavity.  

\subsection{Superradiant phase transition}

\begin{figure}[t]
    \centering
    \includegraphics[width= 0.85\linewidth]{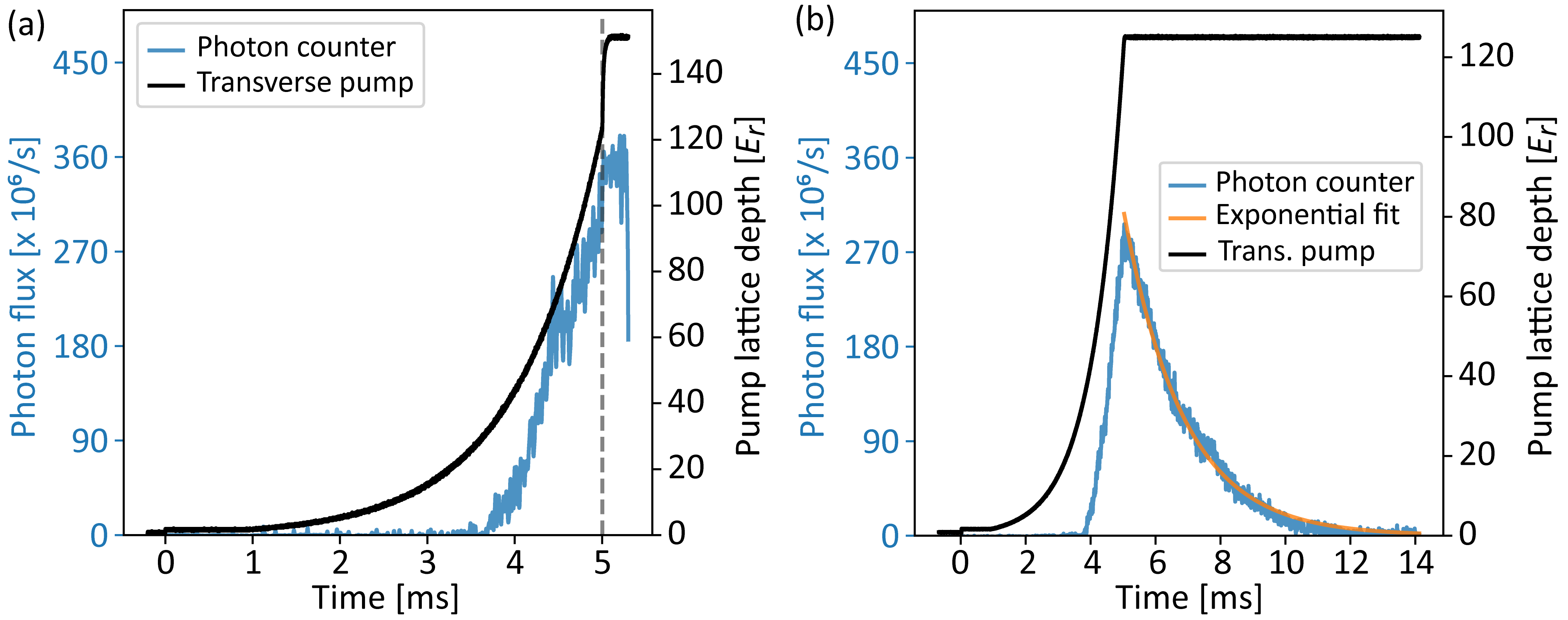}
    \caption{(a) Typical photon-counter trace for an $n=16$ spin glass. The superradiant threshold is crossed close to 4~ms.   The transverse pump is quenched to approximately 150~$E_r$ to maximize photon flux for readout. The quench occurs 5~ms after the start of the ramp, indicated by a dashed line. (b) A photon-counter trace taken with the transverse pump power held constant at $4\Omega_c^2$ exhibits a $1/e$ lifetime of $1.83$~ms in the superradiant phase.}
    \label{fig:SR}
\end{figure}

The transverse-pumping beam is retroreflected to form a 1D standing-wave lattice that is ramped from a lattice depth of zero to approximately 150~$E_r$, as shown in Fig.~\ref{fig:SR}a, where $E_r$ is the recoil energy. This realizes a multimode variant of the Hepp-Lieb-Dicke model~\cite{Kirton2018itt}; see Sec.~\ref{sec:theory} for details. The system exhibits a $\mathbb{Z}_2$ symmetry-breaking transition to a superradiant phase along with the spin glass ordering. This occurs above a critical Rabi frequency of the pump $\Omega_c$ that we estimate to be
\begin{equation}\label{eq:thresh}
    \Omega_c^2 = \frac{2E_r \Delta_A^2 (\Delta_C^2+\kappa^2)}{N_A\eigmax g_0^2|\Delta_C|},
\end{equation}
where $N_A\approx 6\times10^4$ is the number of atoms per ensemble and $\eigmax$ is the largest eigenvalue of the $J$ matrix presented in Sec.~\ref{sec:theory}~\cite{Marsh2024ear}.

Photons emitted from the cavity are detected with a single-photon counter to observe the phase transition and to measure the lifetime in the superradiant phase. Figure~\ref{fig:SR}a shows a typical photon counter trace for our experiments. The phase transition occurs near 4~ms, after which a large number of photons are emitted from the cavity. The photon flux continues to increase over the following 1.5~ms as the transverse pump strength increases. The pump is then quenched to a high value to maximize cavity photon flux for imaging. Figure~\ref{fig:SR}b shows a different experiment in which the transverse pump is held constant at $\Omega^2 = 4\Omega_c^2$ above the transition to measure a lifetime of 1.83~ms in the superradiant phase. The lifetime may be limited by spontaneous emission, molecular recombination, or motional heating of the atoms.

\section{Multimode cavity QED model}\label{sec:theory}
This section presents the derivation of the multimode cavity QED model. It includes a description of the $4/7$ optical resonator and how it supports an effective Ising interaction between spin ensembles trapped within. We derive expressions for the $J$ matrix in Eq.~\eqref{eq:J}, the closely related Green's function of the $4/7$ resonator in Eq.~\eqref{Eq:47Greens}, as well as the emitted cavity field in Eq.~\eqref{eq:modeExp}, which serves as the basis for the fit routines in Sec.~\ref{sec:ImageAnalysis}. 

We begin by considering a network of $n$ spatially separated ultracold gases just below the BEC transition temperature.  These are trapped within a multimode optical cavity and driven with a transverse pump. The total Hamiltonian describing the system is given by $\hat{H}_T=\hat{H}_C+\hat{H}_A+\hat{H}_{LM}$. The first term is the Hamiltonian describing the cavity modes, 
$
    \hat{H}_C=-\hbar\sum_\mu\Delta_\mu \hat{a}_\mu^\dagger \hat{a}_\mu 
$, where the sum over $\mu=(l,m)$ includes all Hermite-Gaussian modes with frequencies $\omega_\mu$ detuned from the transverse pump frequency $\omega_p$ by amounts $\Delta_\mu=\omega_p-\omega_\mu$. The operators $\hat{a}_\mu$ are bosonic lowering operators for each mode. We consider a pump frequency that is near-detuned to a single $\eta$ family of an $M/N$ cavity, for which $\eta=(l+m)\mod N$ is fixed within the family. All modes in other $\eta$ families are far detuned and contribute negligibly. We consider a general $M/N$ cavity before focusing on even-$M$ cavities and then specifically the $4/7$ cavity. The coupled light-matter system is driven-dissipative: The atoms are coherently driven by a transverse pump while the field dissipates from the cavity at a rate $\kappa= 2\pi\cdot 140$~kHz. Dissipation is described by the Lindblad master equation $\dot{\rho}=-i[\hat{H},\rho]/\hbar + \kappa\sum_\mu\mathcal{D}[a_\mu]$, where $\rho$ is the density matrix of the total system and $\mathcal{D}[a]=2a\rho a^\dag - \{a^\dag a,\rho \}$.  The atomic Hamiltonian describes a network of $n$ independent ultracold atomic gases in an external potential~\cite{Pethick2002}:
\begin{equation}\label{eq:Hatomic}
    \hat{H}_{A}=\sum_{i=1}^n \int d^3\mbf{x} \hat{\Psi}_i^\dag(\mbf{x})\bigg(-\frac{\hbar^2{\nabla}^2}{2m}+V_i(\mbf{x}) + \frac{U}{2}\hat{\Psi}^\dag_i(\mbf{x})\hat{\Psi}_i(\mbf{x})
    \bigg)\hat{\Psi}_i(\mbf{x}),
\end{equation}
where $m$ is the atomic mass and the terms $V_i(\mbf{x})$ describe externally applied potentials for trapping each gas.  A bosonic field operator $\hat{\psi}_i(\mbf{x})$ is associated with each gas, where $\mbf{x}=(x,y,z)$ with commutation relations $[\hat{\psi}_i(\mbf{x}),\hat{\psi}_j(\mbf{x}')]=[\hat{\psi}^\dag_i(\mbf{x}),\hat{\psi}^\dag_j(\mbf{x}')]=0$ and $[\hat{\psi}_i(\mbf{x}),\hat{\psi}^\dag_j(\mbf{x}')]=\delta_{ij}\delta(\mbf{x}-\mbf{x}')$.  The term proportional to $U=4\pi\hbar^2a/m$ describes $s$-wave scattering within each gas, where $a\approx 100\cdot a_0$ is the $s$-wave scattering length in terms of the Bohr radius $a_0$~\cite{Pethick2002}. The light-matter coupling Hamiltonian is given by the following expression after elimination of the atomic excited state~\cite{Guo2019eab},
\begin{equation}\label{eq:Hlightmatter}
    \hat{H}_{LM}=\frac{1}{\Delta_A}\sum_{i=1}^n \int d^3\mbf{x} \hat{\Psi}_i^\dag(\mbf{x})\hat{\Phi}_T^\dag(\mbf{x})\hat{\Phi}_T^{}(\mbf{x})\hat{\Psi}_i^{}(\mbf{x}),
\end{equation}
where $\Delta_A=\omega_p-\omega_A=-2\pi\cdot97.2$~GHz is the pump detuning from the bare atomic resonance at $\omega_A$ and
$\hat{\Phi}_T(\mbf{x})\equiv\Omega\cos(k_rx)+g_0\hat{\Phi}(\mbf{x})$. The first term is the standing-wave transverse pump, where $\Omega$ is the Rabi frequency and $k_r=2\pi/\lambda$ is the recoil momentum, and $\lambda=780$~nm is the light wavelength. The second term contains the single-mode, single-photon interaction strength $g_0=2\pi\cdot 1.35$~MHz and the total cavity field operator $\hat{\Phi}(\mbf{x})=\sum_\mu \hat{a}_\mu \Phi_\mu(\mbf{x})$, where $\Phi_\mu(\mbf{x})$ is the 3D mode function~\cite{Guo2019eab,Siegman1986l}
\begin{equation}\label{eq:cavfield_modes}
    \Phi_\mu(\mbf{x}) = \frac{w_0}{w(z)}  \Xi_\mu\bigg(\frac{\mbf{r}}{w(z)}\bigg)\cos\left[k_r\left(z+\frac{\mbf{r}^2}{2R(z)}\right)-\theta_\mu(z)\right].
\end{equation}
Above, $\mbf{r}=(x,y)$ are the coordinates in the transverse plane of the cavity, and $\Xi_\mu(\mbf{r})$ are the Hermite-Gaussian mode profiles. The function $w(z)=w_0\sqrt{1+z^2/z_R^2}$ is the Gaussian spot size where $w_0=34.8$~$\mu$m is the waist of the fundamental mode and $z_R=\pi w_0^2/\lambda=4.87$~mm is the Rayleigh range. $R(z)=z(1+z_R^2/z^2)$ is the wavefront curvature and $\theta_\mu(z)$ is the Gouy phase, given by
\begin{equation}\label{eq:TEMphases}
    \theta_\mu(z)=(1+n_\mu)\psi(z) +n_\mu\psi(L/2)-\xi_\mu,
\end{equation}
where $n_\mu=l+m$, $L$ is the cavity length, and $\psi(z)=\arctan(z/z_R)$. The phase offsets are fixed by the boundary conditions of the cavity and are given by $\xi_\mu=(Q+1)\pi/2+n_\mu\psi(L/2)$, where $Q$ is the longitudinal mode number. This can be simplified using the relation $\psi(L/2)=M\pi/(2N)$~\cite{Guo2019eab} combined with the form of the longitudinal and transverse mode indices in an $\eta$ family: $(Q,l+m)=(Q_0-Mi,\eta+Ni)$ where $Q_0$ is the longitudinal mode number of the lowest-order transverse mode in the $\eta$ family and $i\geq0$ is an integer. These relations yield a simplified expression $\xi_\mu=(1+Q_0+\eta M/N )\pi/2$, showing that the phase offsets depend on only $Q_0$ and $\eta$.

Simplifying the full description of the multimode cavity QED system to an effective spin model requires isolating the relevant degrees of freedom for the atomic fields. The atoms organize in response to the emergent optical potential described by Eq.~\eqref{eq:cavfield_modes}; the transverse pump light interferes with cavity light (arising from photons scattered by the pump) to form a ``checkerboard" potential. The form of the potential that arises in a 4/7 cavity differs from that of the confocal cavity, and we will show how this difference leads to effective Ising, rather than $U(1)$, spin interactions for the 4/7 versus the confocal cavity.

The remainder of this section is organized as follows. Section~\ref{sec:Greens} expands upon the results of Ref.~\cite{Guo2019eab} to derive the Green's function and cavity-mediated interaction of general $M/N$ cavities. This leads to a description of the optical potential experienced by the atoms. In Sec.~\ref{sec:Dicke}, we then isolate the relevant atomic degrees of freedom and map the light-matter system to a multimode Dicke model. In Sec.~\ref{sec:TFieldIsing} the cavity photons are adiabatically eliminated in the dispersive coupling limit. This yields a master equation  solely for the spins that describes a transverse-field Ising model with frustrated interactions. Section~\ref{sec:emittedField} presents the form of the emitted cavity field. 

\subsection{Cavity Green's function}\label{sec:Greens}

The Green's function describes how the phase and amplitude of light propagate between different points in the cavity. This may be calculated analytically for any degenerate or near-degenerate $M/N$ cavity~\cite{Guo2019eab}. The general form of the three-dimensional Green's function $G^\eta(\mbf{x},\mbf{x}')$ for a given $\eta$ family of an $M/N$ cavity is given by
\begin{equation}\label{eq:G3Dbasic}
    G_{3D}(\mbf{x},\mbf{x}',\varphi) = \sum_\mu \Phi_\mu(\mbf{x})\Phi_\mu(\mbf{x}')e^{-n_\mu\varphi}S^\eta_\mu.
\end{equation}
The symbol $S_\mu^\eta$ is a selector defined as $S_\mu^\eta=1$ for modes with $n_\mu=\eta\mod N$ and is zero otherwise. This yields a projection onto the modes within only the $\eta$ family. An exponential mode cutoff is controlled by a parameter $\varphi\geq0$ that may be used to model a finite number of modes and to regularize divergent terms in the Green's function~\cite{Kroeze2023hcu}. An explicit form of the selector function is
\begin{equation}
    S_\mu^\eta = \frac{1}{N}\sum_{s=0}^{N-1}\exp\big[ 2\pi i s (n_\mu - \eta)/N\big].
\end{equation}
By inserting this into the Green's function and expanding the 3D mode functions using Eq.~\eqref{eq:cavfield_modes}, Eq.~\eqref{eq:G3Dbasic} becomes
\begin{equation}\label{eq:G3D_expanded}
\begin{split}
    G_{3D}(\mbf{x},\mbf{x}',\varphi) =  \frac{w_0^2}{Nw(z)w(z')}\sum_{s=0}^{N-1}\sum_\mu &\Xi_\mu\big(\tfrac{\mbf{r}}{w(z)}\big)\Xi_\mu\big(\tfrac{\mbf{r}'}{w(z')}\big)\exp\big[-n_\mu\varphi + 2\pi i s (n_\mu - \eta)/N\big]\\ &\quad\times\cos\bigg[k_r\left(z+\frac{\mbf{r}^2}{2R(z)}\right)-\theta_\mu(z)\bigg]\cos\bigg[k_r\left(z'+\frac{\mbf{r'}^2}{2R(z')}\right)-\theta_\mu(z')\bigg].
\end{split}
\end{equation}
The summation over cavity modes is now performed explicitly using the Mehler kernel, the Green's function for the quantum harmonic oscillator, which is 
\begin{equation}\label{eq:Mehler}
    G(\mbf{r},\mbf{r}',\phi) = \sum_{\mu}\Xi_\mu(\mbf{r})\Xi_\mu(\mbf{r}')e^{-n_\mu \phi}= \frac{e^\phi}{2\pi\sinh(\phi)}\exp\left[ -\frac{(\mbf{r}-\mbf{r}')^2}{2\tanh(\phi/2)} -\frac{(\mbf{r}+\mbf{r}')^2}{2\coth(\phi/2)} \right].
\end{equation}
The sum is taken over all Hermite-Gauss modes and the parameter $\phi$ may be any complex number with nonnegative real part. We define a modified Mehler kernel that incorporates the selector function and normalization by $w(z)$ as
\begin{equation}\label{eq:modGreen}
    G^\eta(\mbf{r},\mbf{r}',\varphi) = \sum_{\mu}\Xi_\mu\big(\tfrac{\mbf{r}}{w(z)}\big)\Xi_\mu\big(\tfrac{\mbf{r}'}{w(z')}\big)e^{-n_\mu \varphi} S_\mu^\eta = \frac{1}{N}\sum_{s=0}^{N-1}e^{-\eta 2\pi i s/N}G\big(\tfrac{\mbf{r}}{w(z)},\tfrac{\mbf{r}'}{w(z')},\varphi-2\pi is/N\big). 
\end{equation}

The 3D Green's function $G_{3D}$ can now be evaluated in terms of this modified Green's function by expanding the cosine functions in Eq.~\eqref{eq:G3D_expanded} using complex exponentials. This yields a matrix form of the Green's function:
\begin{equation}\label{eq:G3Dmatrix}
    G_{3D}(\mbf{x},\mbf{x}',\varphi) =\begin{pmatrix} \cos\zeta \\ \sin\zeta \end{pmatrix}\cdot{D}_{3D}(\mbf{x},\mbf{x}',\varphi)\begin{pmatrix} \cos\zeta' \\ \sin\zeta' \end{pmatrix},
\end{equation}
where we defined coordinates via $\zeta = k_r[z + \mbf{r}^2/2R(z)]-\psi(z)$ and define $D_{3D}$ as a $2\times 2$ interaction matrix.  It takes the following general form for any $\varphi\geq0$:
\begin{equation}\label{eq:D3Dgeneral}
\begin{split}
    {D}_{3D}(\mbf{x},\mbf{x}',\varphi) =\frac{w_0^2}{2w(z)w(z')}\Re &\bigg[\big( \mbf{1} + \sigma_y  \big) G^\eta\big(\mbf{r},\mbf{r}',\varphi+i\psi(z)-i\psi(z')\big)\\&\,\,- e^{i\pi Q_0}\big( \sigma_z +i\sigma_x \big) e^{i\eta M\pi/N} G^\eta\big(\mbf{r},\mbf{r}',\varphi+i\psi(z)+i\psi(z')+iM\pi/N\big)  \bigg],
\end{split}
\end{equation}
where $\sigma_{x/y/z}$ are Pauli matrices. We now describe the significance of the terms in Eqs.~\eqref{eq:G3Dmatrix} and~\eqref{eq:D3Dgeneral}. The components $\cos(\zeta)$ and $\sin(\zeta)$ are the longitudinal quadratures of the cavity field, and the matrix $D_{3D}$ describes how a source field of a given quadrature at location $\mbf{x'}$ propagates to a different location $\mbf{x}$ while evolving in phase. In other words, $D_{3D}$ describes how a $\cos(\zeta')$ source field may create a $\cos(\zeta)$ or $\sin(\zeta)$ field elsewhere in the cavity. The diagonal matrices $\mbf{1}$ and $\sigma_z$ above describe components of the Green's function that retain the phase of the source, while the off-diagonal matrices $\sigma_{x/y}$ describe mixing of longitudinal modes. From the above, we see that the $\sigma_y$ interaction drops out whenever $z=z'$ because $\sigma_y$ is purely imaginary.

A simplification is possible when specifically considering the midplane of the cavity, $z=z'=0$, which is where we trap the atomic ensembles. At the midplane, $w(z)=w_0$, $1/R(z)=0$, and $\psi(z)=0$. The interaction matrix at this location is $D(\mbf{r},\mbf{r}',\varphi)={D}_{3D}(\mbf{r},z=0,\mbf{r}',z'=0,\varphi)$ for a general $M/N$ cavity and simplifies to
\begin{equation}\label{eq:midplaneGreen}
    {D}(\mbf{r},\mbf{r}',\varphi) =\frac{1}{2} \mbf{1} G^\eta\big(\mbf{r},\mbf{r}',\varphi\big)- \frac{(-1)^{Q_0}}{2}\Re\bigg[(\sigma_z+i\sigma_x) e^{i\eta M\pi/N} G^\eta\big(\mbf{r},\mbf{r}',\varphi+iM\pi/N\big)  \bigg].
\end{equation}
Further simplification of $D_{3D}$ is possible for  $M/N$ cavities with even $M$, such as the 4/7 cavity.  Starting with the following relation for any integer $l$: 
\begin{equation}\label{eq:evenMidentity}
    G^\eta\big(\mbf{r},\mbf{r}',\varphi+ilM\pi/N\big) = e^{-i\eta l M\pi/N} G^\eta\big(\mbf{r},\mbf{r}',\varphi\big),
\end{equation}
we see that this relation allows $D_{3D}$ to be simplified for even-$M$ cavities to 
\begin{equation}
\begin{split}
    {D}_{3D}(\mbf{x},\mbf{x}',\varphi) =\frac{w_0^2}{2w(z)w(z')}\Re &\bigg[\big( \mbf{1} + \sigma_y  \big) G^\eta\big(\mbf{r},\mbf{r}',\varphi+i\psi(z)-i\psi(z')\big)\\&\,\,- e^{i\pi Q_0}\big( \sigma_z +i\sigma_x \big) G^\eta\big(\mbf{r},\mbf{r}',\varphi+i\psi(z)+i\psi(z')\big)  \bigg].
\end{split}
\end{equation}
Moreover, the midplane interaction matrix for an even-$M$ cavity reduces to 
\begin{equation}
    {D}(\mbf{r},\mbf{r}',\varphi) =\frac{1}{2} \big(\mbf{1} - e^{i\pi Q_0}\sigma_z  \big)G^\eta\big(\mbf{r},\mbf{r}',\varphi\big) 
    =\begin{cases}
        \begin{pmatrix}
            1 & 0 \\
            0 & 0
        \end{pmatrix}G^\eta\big(\mbf{r},\mbf{r}',\varphi\big) & Q_0\,\,\mathrm{odd} \\
        \begin{pmatrix}
            0 & 0 \\
            0 & 1
        \end{pmatrix}G^\eta\big(\mbf{r},\mbf{r}',\varphi\big) & Q_0\,\,\mathrm{even}.
    \end{cases}
\end{equation}
We see that one of the longitudinal quadratures of the cavity completely drops out regardless of whether $Q_0$ is even or odd:  cavity modes of only a single longitudinal quadrature are supported in the midplane of even-$M$ cavities. This fact is the primary result of this section. It implies that the emergent optical potential experienced by atoms in the midplane can have only a single longitudinal form, as opposed to the odd-$M$ confocal cavity for which the phase of the cavity field can vary continuously~\cite{Guo2021aol}.

The Green's function at the midplane of an even-$M$ cavity is thus fully described by $Q_0$ and the modified Green's function $G^\eta$ in Eq.~\eqref{eq:modGreen}. We now provide the explicit form of $G^\eta$ for the $4/7$ cavity in the $\varphi=0$ limit of perfect mode degeneracy:
\begin{equation}
\label{Eq:47Greens}
    G^\eta\big(\mbf{r},\mbf{r}',\varphi=0\big) = \frac{1}{7}\delta\left(\frac{\mbf{r}-\mbf{r}'}{w_0}\right) +   \frac{1}{7\pi}\sum_{k=1}^3\frac{1}{\sin(2k\pi/7)} \sin\left[ (1+\eta)\frac{2k\pi}{7} + \frac{\mbf{r}^2 + {\mbf{r}'}^2}{\tan(2k\pi/7)w_0^2} - \frac{2 \mbf{r}\cdot\mbf{r}'}{\sin(2k\pi/7)w_0^2}\right]. 
\end{equation}
The delta function term indicates a strong local field forming at the source location, indicating constructive interference of the modes in the $\eta$ family at that location. 
Propagation of the local field over $N=7$ round trips in the cavity produces three additional terms. These correspond to defocused components of the electric field that are nonlocal and sign-changing, similar to the $\cos(2\mbf{r}\cdot\mbf{r}'/w_0^2)$ nonlocal field of the confocal cavity~\cite{Guo2019spa}. As presented in Eq.~\eqref{Eq:47Greens}, the Green's function can still be calculated analytically for $\varphi>0$ using Eq.~\eqref{eq:modGreen}.  However, it no longer assumes a simple analytic form like for the confocal case. Qualitatively, $\varphi$ regularizes the delta function to yield local interactions that are of finite strength and spatial width. The nonlocal fields are less impacted, but experience a diminishment of the components of the field that have high spatial frequency. We estimate $\varphi<6\times10^{-4}$ for our cavity~\cite{Kroeze2023hcu}, making the effect of finite mode support negligible compared to the effect of finite spatial extent of the atomic ensembles; this is discussed in Sec.~\ref{sec:ImageAnalysis}. 

\subsection{Mapping to the multimode Dicke model}\label{sec:Dicke}
We now map the multimode cavity QED system onto a multimode Dicke model.  This uses the single-quadrature form of the emergent optical potential in the previous subsection. The Dicke model is realized by mapping atomic momentum states onto effective spin degrees of freedom. Each spin is represented as the phase and amplitude of an atomic density wave that forms in response to the intracavity 2D lattice.  This lattices arises from the interference between transverse pump and emergent intracavity light. A ``checkerboard" density wave minimizes atomic energy within the lattice. To describe this degree of freedom, we expand the atomic field in terms of Fourier components as 
\begin{equation}\label{eq:ansatz}
    \hat{\Psi}_i(\mbf{x}) = \sqrt{\rho_i(\mbf{x})}\left[\hat{\psi}_{0,i} + 2\hat{\psi}_{c,i}\cos(k_rz)\cos(k_rx)\right],
\end{equation}
where $\rho_i(\mbf{x})$ is a normalized envelope function that varies slowly over $\hat{z}$~\cite{Guo2019eab} and is centered at $\mbf{x}_i$, the trap center for each gas. The operators $\hat{\psi}_{0,i}$ and $\hat{\psi}_{0,i}$ are bosonic lowering operators for the background and density-wave components of the gas, resp. This two-level approximation ignores finite-temperature effects in the atomic ensemble; this approximation has little effect on the resulting Hepp-Lieb-Dicke model except for a shift in the superradiant threshold~\cite{Piazza2013bcv}.  Unlike in our previous work~\cite{Kroeze2023rsb}, the ansatz contains no component $\propto \sin(k_rz)$ because the $4/7$ cavity we employ now  supports modes of only a single longitudinal form at the midplane, as described above. We consider an odd $Q_0$ so that only the $\cos(k_rz)$ modes are supported. The choice of even $Q_0$ leads to the same spin model. Inserting the density-wave ansatz results in the atomic Hamiltonian 
\begin{equation}
    \hat{H}_{A,i} = 2E_r\hat{\psi}_{c,i}^\dag \hat{\psi}_{c,i}^{}  + E_{\text{trap},i}\big(\hat{\psi}_{0,i}^\dag \hat{\psi}_{0,i}^{}+\hat{\psi}_{c,i}^\dag \hat{\psi}_{c,i}^{} \big) ,
\end{equation}
where $E_r=\hbar^2k_r^2/(2m)$ is the recoil energy. The trap energy $E_{\text{trap},i}=\int d^3\mbf{x}V(\mbf{x})\rho(\mbf{x})$ is the same for both the $\hat{\psi}_{0}$ and $\hat{\psi}_{c}$ states and as such, results in a constant energy offset that can be ignored. Insertion of the density-wave ansatz into the light-matter coupling Hamiltonian yields
\begin{equation}\begin{split}
    \frac{\hat{H}_{LM,i} }{\hbar}= & \frac{\Omega^2}{2\Delta_A}\Big(\hat{\psi}_{0,i}^\dag \hat{\psi}_{0,i}^{} +\frac{3}{2}\hat{\psi}_{c,i}^\dag \hat{\psi}_{c,i}^{}\Big) +\frac{g_0\Omega}{2\Delta_A}\sum_\mu \cos\theta_\mu S_\mu^\eta (\hat{a}_\mu+\hat{a}_\mu^\dag) \big(\hat{\psi}_{0,i}^{}\hat{\psi}_{c,i}^\dag + \hat{\psi}_{0,i}^\dag\hat{\psi}_{c,i}\big)\int d^2\mbf{r}\rho_i(\mbf{r})\Xi_\mu(\mbf{r}),
\end{split}\end{equation}
where $\rho_i(\mbf{r})=\int dz \rho_i(x,y,z)$ is the atomic density profile in the transverse plane. We ignore the dispersive shift terms $\propto g_0^2/\Delta_A$ because they play but a little role. Mapping to a spin model is achieved using $SU(2)$ collective spin operators given by $\hat{S}_i^z = (\hat{\psi}_{c,i}^\dag \hat{\psi}_{c,i}^{} - \hat{\psi}_{0,i}^\dag \hat{\psi}_{0,i}^{})/2$ and $\hat{S}_i^x=(\hat{\psi}_{0,i}^{}\hat{\psi}_{c,i}^\dag + \hat{\psi}_{0,i}^\dag\hat{\psi}_{c,i})/2$. Writing the Hamiltonian in terms of spin operators and ignoring constant energy offsets yields a multimode Dicke model,
\begin{equation}\label{eq:mmDicke}
    \frac{\hat{H}_\mathrm{Dicke}}{\hbar} = -\sum_\mu\Delta_\mu \hat{a}_\mu^\dagger \hat{a}_\mu  +\omega_z\sum_{i=1}^n \hat{S}_i^z + \sum_\mu\sum_{i=1}^ng_{i\mu} (\hat{a}_\mu+\hat{a}_\mu^\dag) \hat{S}_i^x,
\end{equation}
where $\omega_z=2E_r/\hbar+\Omega^2/(4\Delta_A)$ and the effective light-matter coupling strengths are given by
\begin{equation}
    g_{i\mu} = \cos\theta_\mu S_\mu^\eta\frac{g_0\Omega}{\Delta_A}\int d^2\mbf{r}\rho_i(\mbf{r})\Xi_\mu(\mbf{r}).
\end{equation}

\subsection{Transverse-field Ising spin model}\label{sec:TFieldIsing}
The cavity modes can be adiabatically eliminated to yield a quantum description that depends only on the spin degrees of freedom. This approximation is accurate when the cavity detunings $\Delta_\mu$ are much larger in magnitude than the atomic frequency $\omega_z$ and cavity decay rate $\kappa$. We previously derived in Ref.~\cite{Marsh2024ear} the atom-only model for the multimode Dicke model in Eq.~\eqref{eq:mmDicke} using the method of J\"ager et al.~\cite{Jager2022lme}. In the limit $|\Delta_\mu|/\kappa\gg 1$, the atom-only Hamiltonian is given by 
\begin{equation}
    \frac{\hat{H}}{\hbar} = \omega_z \sum_{i=1}^n \hat{S}_i^z - \frac{g_0^2\Omega^2}{\Delta_A^2|\Delta_C|}\sum_{ij=1}^n J_{ij}\hat{S}_i^x\hat{S}_j^x,
\end{equation}
where $\Delta_C$ is the detuning of the fundamental mode. The above corresponds to Eq.~\eqref{eq:Ham} where we ignored the small dispersive shift $\propto\Omega^2/\Delta_A$ of the atomic frequency in the main text. The term $J_{ij}$ is the dimensionless cavity-mediated interaction matrix,
\begin{equation}\label{eq:J}
\begin{split}
    J_{ij} &= \Delta_C\int d\mbf{r} d\mbf{r}' \rho_i(\mbf{r})\rho_j(\mbf{r}')\sum_\mu \frac{S_\mu^\eta\Delta_\mu}{\Delta_\mu^2+\kappa^2}  \Xi_\mu(\mbf{r})\Xi_\mu(\mbf{r}') \\
    &= \int d\mbf{r} d\mbf{r}' \rho_i(\mbf{r})\rho_j(\mbf{r}') G^\eta(\mbf{r},\mbf{r}',\varphi),
\end{split}
\end{equation}
where $G^\eta$ is the cavity Green's function given by Eq.~\eqref{eq:modGreen}, for general $\varphi$, and specifically Eq.~\eqref{Eq:47Greens} in the $\varphi=0$ limit. In writing the above equation, we approximate the mode dispersion $\Delta_C\Delta_\mu/(\Delta_\mu^2+\kappa^2)$ by the exponential form $\exp(-n_\mu\varphi)$ for small $\varphi$. This captures the finite resolution of the cavity field.
We find in Sec.~\ref{sec:ImageAnalysis} that the experimental data fits well to $\varphi=0$ when the atomic densities $\rho_i(\mbf{x})$ are faithfully accounted for. Thus, we  approximate $\varphi$ as equal to 0 in our analyses while explicitly accounting for the finite spatial extent of $\rho_i(\mbf{r})$, as described in Sec.~\ref{sec:ImageAnalysis}. 

A Lindblad master equation accounts for the dissipation of cavity modes. This was derived in the atom-only model of Ref.~\cite{Marsh2024ear} to be
$
    \dot{\rho}= -i[\hat{H},\rho]/\hbar + \sum_{i=1}^n \mathcal{D}[\hat{C}_i],
$
where the collapse operators $\hat{C}_i$ are given by 
\begin{equation}
    \hat{C}_i = \frac{\sqrt{\lambda_i\kappa}g_0\Omega}{2|\Delta_C|\Delta_A}\sum_{j=1}^n \mbf{v}_j^i \hat{S}_j^x.
\end{equation}
Here, $\lambda_i$ and $\mbf{v}^i$ are the $i$'th eigenvalue and eigenvector of the semipositive-definite $J$ matrix, resp. The total dissipation rate \emph{per spin} at threshold is $\sum_i^n\ex{ \hat{C}_i^\dag \hat{C}_i}/n\approx N_A\kappa\omega_z/|\Delta_C|$~\cite{Marsh2024ear}, where $N_A\approx6\cdot 10^4$ is the number of atoms in each spin ensemble. The dissipation rate per spin is $2\pi\cdot 3$~MHz in the experimental parameter regime explored.

\subsection{Emitted cavity field}\label{sec:emittedField}
We now describe how the transmitted cavity field depends on the spin state, enabling spin-state detection through holographic imaging~\cite{Guo2019spa}. The emitted field is determined by computing the steady-state expression for the cavity field. This is performed by setting the time derivative of the cavity mode operators to zero in the Heisenberg picture.   Using the Lindblad equation corresponding to the total Hamiltonian $\hat{H}_T$ and $\kappa$ yields
\begin{equation}\label{eq:modeExp}
    \hat{a}_\mu = \sum_{i=1}^n \frac{g_{i\mu}\hat{S}_i^x}{\Delta_\mu+i\kappa}.
\end{equation}
This steady state expression is now inserted into  $\hat{\Phi}(\mbf{x})=\sum_\mu \hat{a}_\mu \Phi_\mu(\mbf{x})$ to obtain the steady-state cavity field. The total field $\hat{\Phi}(\mbf{x})$ is a standing wave ${\propto}\cos(k_rz)$ composed of forward and backward propagating fields $\hat{\Phi}(\mbf{x})=(\hat{\Phi}_F(\mbf{x})e^{-ik_rz} + \hat{\Phi}_B(\mbf{x})e^{ik_rz})/2$. The forward-propagating field $\hat{\Phi}_F(\mbf{x})$ is the component that is detected with phase-sensitive holographic imaging~\cite{Guo2019spa}. Inserting the steady-state cavity field and taking the expectation value of the forward-propagating component results in
\begin{equation}\label{eq:emitted}
    \ex{\hat{\Phi}_F(\mbf{x})} = \frac{g_0\Omega}{\Delta_A\Delta_C}\sum_{i=1}^n \ex{\hat{S}_i^x}\int d^2\mbf{r}' \rho_i(\mbf{r}') G^\eta(\mbf{r},\mbf{r}').
\end{equation}
We ignore a small term ${\propto} \kappa/\Delta_C$ that is negligible in the experimental parameter regime. The above expression forms the basis for the fitting routine used to extract the spin components $\ex{\hat{S}_i^x}$, as described in Sec.~\ref{sec:ImageAnalysis}.

\section{Image analysis}\label{sec:ImageAnalysis}

\begin{figure}[t!]
    \centering
    \includegraphics[width= \linewidth]{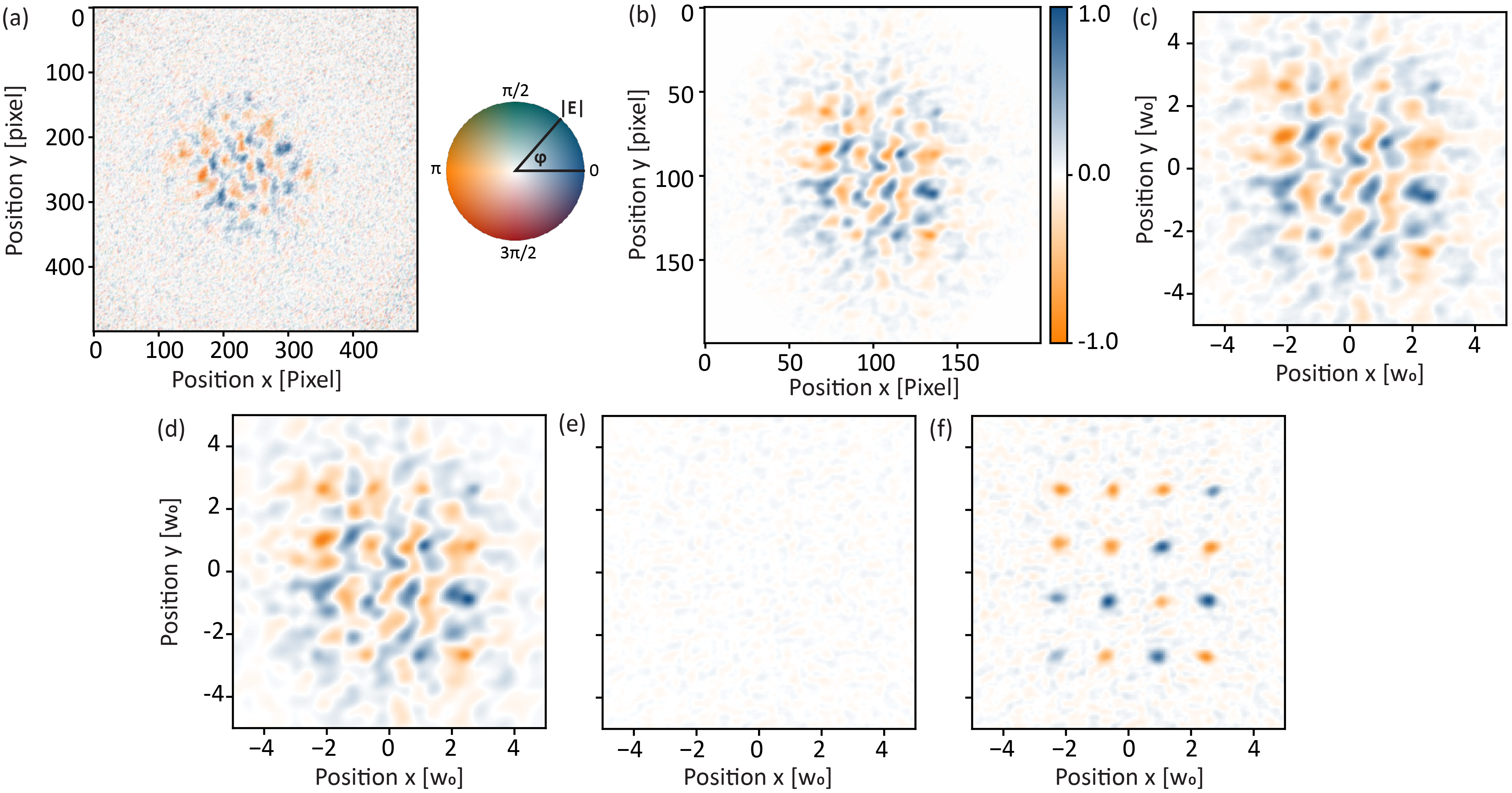}
    \caption{ Image analysis overview. (a) An example of a holographic image of the emitted cavity field. The color wheel indicates the phase and amplitude of the complex field. (b) Image after rotation, downsampling, and removal of imaginary component.  The color bar indicates the field amplitude after normalization by the maximum absolute pixel value. (c) The image is filtered using $\tilde{\mathcal{F}}$ to remove noise and to extract the cavity center and waist $w_0$. (d) Example of the least-squares fit to the image in panel c. (e) Fit residual. (f)  Spin distributions within each ensemble are estimated by subtracting the nonlocal components of the fit; see text for details.}
    \label{fig:imageAnalysis}
\end{figure}

Phase-sensitive imaging of the electric field at the cavity midplane is performed using holographic reconstruction from cavity emission. This enables microscopic spin-state detection. Experimental implementation of holographic imaging is described in our previous publications~\cite{Guo2019spa,Kroeze2023rsb}. An example holographic image of $n=16$ spins for disorder realization $J_1$ is shown Fig.~\ref{fig:imageAnalysis}a. The color and intensity of the image map onto the phase and amplitude of the electric field, respectively, as indicated by the color wheel.  Section~\ref{sec:theory} discussed how the mode structure of the 4/7 cavity limits the phase of the electric field in the midplane to 0 or $\pi$. Indeed, we observe that after a global phase rotation, the measured 4/7 field is predominantly 0 or $\pi$ in phase except where the signal level approaches the noise floor. In Fig.~\ref{fig:imageAnalysis}b, the image is rotated to align with the XODT beam directions and downsampled by a factor of 2 in each direction to reduce the image size. Down sampling reduces the time required for fitting, enabling the position feedback described in Sec.~\ref{ColdAtomProd} while introducing a negligible effect on the extraction of spin amplitudes. The imaginary component of the image that remains is due to technical noise and discarded. We now describe the additional processing and fit routine used to analyze the images and extract spin states.

\subsection{Fractional Fourier transform analysis}

$M/N$ cavities display symmetries under particular fractional Fourier transforms. We utilize these symmetries to extract the cavity center coordinate from the images, to calibrate the size of the fundamental mode waist $w_0$ in terms of camera pixels, and to filter the images to remove noise. The symmetry is derived starting from the paraxial description of the half-round-trip propagator of the cavity. This is calculated in terms of an ABCD matrix $M_\text{Half-trip}$ that describes propagation from the cavity midplane to one of the mirrors, reflection, and propagation back to the midplane~\cite{Siegman1986l}, 
\begin{equation}
  M_\text{Half-trip}=\begin{pmatrix}
        1 & L/2 \\
        0 & 1
    \end{pmatrix}
    \begin{pmatrix}
        1 & 0 \\
        -2/R & 1
    \end{pmatrix}
    \begin{pmatrix}
        1 & L/2 \\
        0 & 1
    \end{pmatrix},\\
\end{equation}
where $L$ is the cavity length and $R$ the mirror radius of curvature. By making use of the multimode condition $L/R=2\sin^2[M\pi/(2N)]$~\cite{Guo2019eab} and a series of algebraic manipulations, $M_\text{Half-trip}$ can be written as
\begin{equation}\label{eq:HalfTrip}
  M_\text{Half-trip}=\begin{pmatrix}
        z_R^{-1} & 0  \\
        0 & 1
    \end{pmatrix}
    \begin{pmatrix}
        \cos(M\pi/N) & \sin(M\pi/N) \\
        -\sin(M\pi/N) & \cos(M\pi/N)
    \end{pmatrix}
    \begin{pmatrix}
        z_R & 0 \\
        0 & 1
    \end{pmatrix}=\begin{pmatrix}
        \cos(M\pi/N) & z_R\sin(M\pi/N) \\
        -z_R^{-1}\sin(M\pi/N) & \cos(M\pi/N)
    \end{pmatrix},
\end{equation}
where $z_R=\pi w_0^2/\lambda$ is the Rayleigh range. Thus, $M_\text{Half-trip}$ is a rotation matrix that generates a rotation by an angle $-M\pi/N$ up to a rescaling of the $z$ coordinate by $z_R$.

The connection to the fractional Fourier transform is revealed by writing $M_\text{Half-trip}$ in terms of an explicit propagator for the electric field. The relationship between a general ABCD transfer matrix $\big(\begin{smallmatrix}A & B\\ C & D\end{smallmatrix})$ and its propagator for the electric field is described by the kernel~\cite{Siegman1986l}
\begin{equation}\label{eq:prop}
    K_{ABCD}(\mbf{r}_2,\mbf{r}_1) = \frac{i}{ B\lambda}\exp[-\frac{i\pi}{B\lambda}(A\mbf{r}_1^2 - 2 \mbf{r}_1 \cdot \mbf{r}_2 + D\mbf{r}_2^2)].
\end{equation}
The element $C$ does not appear; it has been eliminated by the constraint $\det(M)=AD-BC=1$ for optical systems in free space~\cite{Siegman1986l}. $K_{ABCD}$ is used to propagate an electric field $E_1(\mbf{r})$ under the action of the ABCD matrix to form a new field $E_2(\mbf{r})$ given by
\begin{equation}
    E_2(\mbf{r}) = e^{-i2\pi L_0/\lambda}\iint_{-\infty}^\infty  K_{ABCD}(\mbf{r},\mbf{r}')E_1(\mbf{r}') d\mbf{r}',
\end{equation}
where $L_0$ is the total optical path length of the ABCD optic~\cite{Siegman1986l}, which is assumed to be cylindrically symmetric. $L_0=L$ for $M_\text{Half-trip}$ such that $L_0/\lambda$ is an integer and the phase factor is equal to unity. In the case of $M_\text{Half-trip}$ the propagator takes the following form by inserting the elements of the ABCD matrix in Eq.~\eqref{eq:HalfTrip} into the formula of Eq.~\eqref{eq:prop}, yielding
\begin{equation}
    K_\text{Half-trip}(\mbf{r}_2,\mbf{r}_1)=\frac{i}{\pi w_0^2\sin\left(\frac{M\pi}{N}\right)}\exp\left[ -i\cot\left(\tfrac{M\pi}{N}\right)\frac{(\mbf{r}_1^2+\mbf{r}_2^2)}{w_0^2} +2i\csc\left(\tfrac{M\pi}{N}\right)\frac{\mbf{r}_1\cdot\mbf{r}_2}{w_0^2}  \right].
\end{equation}
The propagator $K_\text{Half-trip}$ has a close connection to the fractional Fourier transform (FRFT) $\mathcal{F}_\alpha$. The FRFT is also represented by an integral form $\mathcal{F}_\alpha[E(\mbf{u})](\mbf{v})=\iint K^\mathcal{F}_\alpha(\mbf{v},\mbf{u})E(\mbf{u})d\mbf{u}$, with a kernel that can be written as~\cite{ozaktas2001tff}
\begin{equation}
    K^\mathcal{F}_\alpha(\mbf{u},\mbf{v})=\frac{(1+i\cot(\alpha))}{\pi}\exp\Big[-i\cot(\alpha)(\mbf{u}^2+\mbf{v}^2)+2i\csc(\alpha)\mbf{u}\cdot\mbf{v} \Big].
\end{equation}
The parameter $\alpha\in[0,2\pi]$ is the angle of the transformation. For convenience, we use the convention that $\alpha=-\pi/2$ corresponds to the standard Fourier transform. Values of $\alpha$ that are not integer multiples of $\pi/2$ correspond to intermediate transformations between Fourier and real space. We find that the half-trip propagator and the FRFT kernel are related by
\begin{equation}\label{eq:HalfTripFRFT}
    K_\text{Half-trip}(\mbf{r}_2,\mbf{r}_1)=\frac{e^{i\frac{M\pi}{N}}}{w_0^2}K^\mathcal{F}_{\frac{M\pi}{N}}\left(\frac{\mbf{r}_2}{w_0},\frac{\mbf{r}_1}{w_0}\right).
\end{equation}
A half-trip propagation in an $M/N$ cavity therefore corresponds to an FRFT of angle $M\pi/N$. The semigroup property of the FRFT means that $l$ half-trip propagations correspond to a single FRFT of angle $lM\pi/N$. $N$ round trips, or $2N$ half trips, thus correspond to a rotation by an angle $2\pi M$. Since this is an integer multiple of $2\pi$, $N$ round trips corresponds to the identity operator and leaves any field invariant.  

Any steady-state cavity field $E(\mbf{r}/w_0)$ is, by definition, invariant under propagation in the cavity. Therefore, applying any FRFT of angle $lM\pi/N$ for an integer $l$ leaves the cavity field invariant up to a phase factor given by Eq.~\eqref{eq:HalfTripFRFT}. In particular, we consider a symmetry-averaging operator $\tilde{\mathcal{F}}$ that sums over all $N$ distinct FRFTs of angle $lM\pi/N$:
\begin{equation}
    \tilde{\mathcal{F}}[E](\mbf{r}/w_0)=\frac{1}{N}\sum_{l=0}^N e^{ilM\pi/N}\mathcal{F}_{\frac{lM\pi}{N}}[E](\mbf{r}/w_0).
\end{equation}
While any steady-state cavity $E(\mbf{r}/w_0)$ is invariant under $\tilde{\mathcal{F}}$, noise terms arising from, e.g., camera noise, are not. Therefore, $\tilde{\mathcal{F}}$ acts as a filter that removes any component of the measured images that do not conform to the symmetries of the $M/N$ cavity. Moreover, cavity fields are invariant under $\tilde{\mathcal{F}}$ only when the waist $w_0$ is known in pixel units of the camera, and when the coordinates $\mbf{r}$ are referenced to the center of the cavity. The measured images themselves can therefore be used to calibrate $w_0$ and the cavity center location on the camera. To do so, given an image $I(x,y)$ such as that presented in Fig.~\ref{fig:imageAnalysis}b, we minimize the least-squares cost function
\begin{equation}
    C=\frac{1}{2}\sum_{ij}\left( \tilde{\mathcal{F}}[I]\left(\frac{(x_i-x_c)}{w_0},\frac{(y_j-j_c)}{w_0}\right) -I(x_i,y_j) \right)^2
\end{equation}
over $w_0$, $x_c$, and $y_c$ to extract the waist and cavity center location for each image. These parameters vary little ($<1\%$) between experimental shots, but can exhibit a slow drift that we account for through this minimization procedure. Performing this minimization results in experimental images such as that in Fig.~\ref{fig:imageAnalysis}c, which are symmetry-averaged, translated to the cavity center, and scaled by the waist $w_0$. 

\subsection{Fitting routine}

The symmetry-averaged images are now fit to a model of the cavity field that incorporates the spin degrees of freedom. The fit form is derived from Eq.~\eqref{eq:emitted} for the emitted cavity field,
\begin{equation}
    I(\mbf{r})=\sum_{i=1}^n A_i \sum_{x',y'} \rho_i(\mbf{r}')G^\eta(\mbf{r},\mbf{r}'),
\end{equation}
where the sum over $\mbf{r}'=(x',y')$ traverses all pixels in the image, $A_i$ is an amplitude for each spin ensemble, $G^\eta$ is the cavity Green's function given by Eq.~\eqref{Eq:47Greens}, and each $\rho_i(\mbf{r})$ is an atomic density. We consider atomic densities of the form
\begin{equation}
    \rho_i(\mbf{r})=a^i_{00}\text{HG}_{00}\left(\frac{(x-x_i)}{\sqrt{2}\sigma_x},\frac{(y-y_i)}{\sqrt{2}\sigma_y}\right) + a^i_{01}\text{HG}_{01}\left(\frac{(x-x_i)}{\sqrt{2}\sigma_x},\frac{(y-y_i)}{\sqrt{2}\sigma_y}\right) + a^i_{10}\text{HG}_{10}\left(\frac{(x-x_i)}{\sqrt{2}\sigma_x},\frac{(y-y_i)}{\sqrt{2}\sigma_y}\right).
\end{equation}
The function $\text{HG}_{lm}$ are the Hermite-Gauss mode functions. $\text{HG}_{00}[(x-x_i)/(\sqrt{2}\sigma_x),(y-y_i)/(\sqrt{2}\sigma_y)]$ is thus a 2D Gaussian function of standard deviations $\sigma_{x/y}$ centered at location $\mbf{r}_i=(x_i,y_i)$. This component of the atomic density extracts the fraction of the ensemble that aligns along one direction as an effective spin. The components $\text{HG}_{01}$ and  $\text{HG}_{10}$ are included to allow for the possibility that a fraction of the atomic ensemble anti-aligns with the rest. This can occur when nodes of the total cavity field appear within an ensemble. The atoms then experience a competition between their local interactions, which drive effective spin alignment within the ensemble, and the nonlocal interactions, which drive anti-alignment of the atoms that reside on opposing sides of the field node; see Ref.~\cite{Kollar2017scw} for examples where field nodes were specially positioned within atomic ensembles. We normalize the fit coefficients $a^i_{lm}$ such that $(a^i_{00})^2+(a^i_{01})^2+(a^i_{10})^2=1$ and that $(a^i_{lm})^2$ represents the fraction of spins organized in the $\text{HG}_{lm}$ configuration. 

\begin{figure}[t]
\centering
\includegraphics[width=0.85\textwidth]{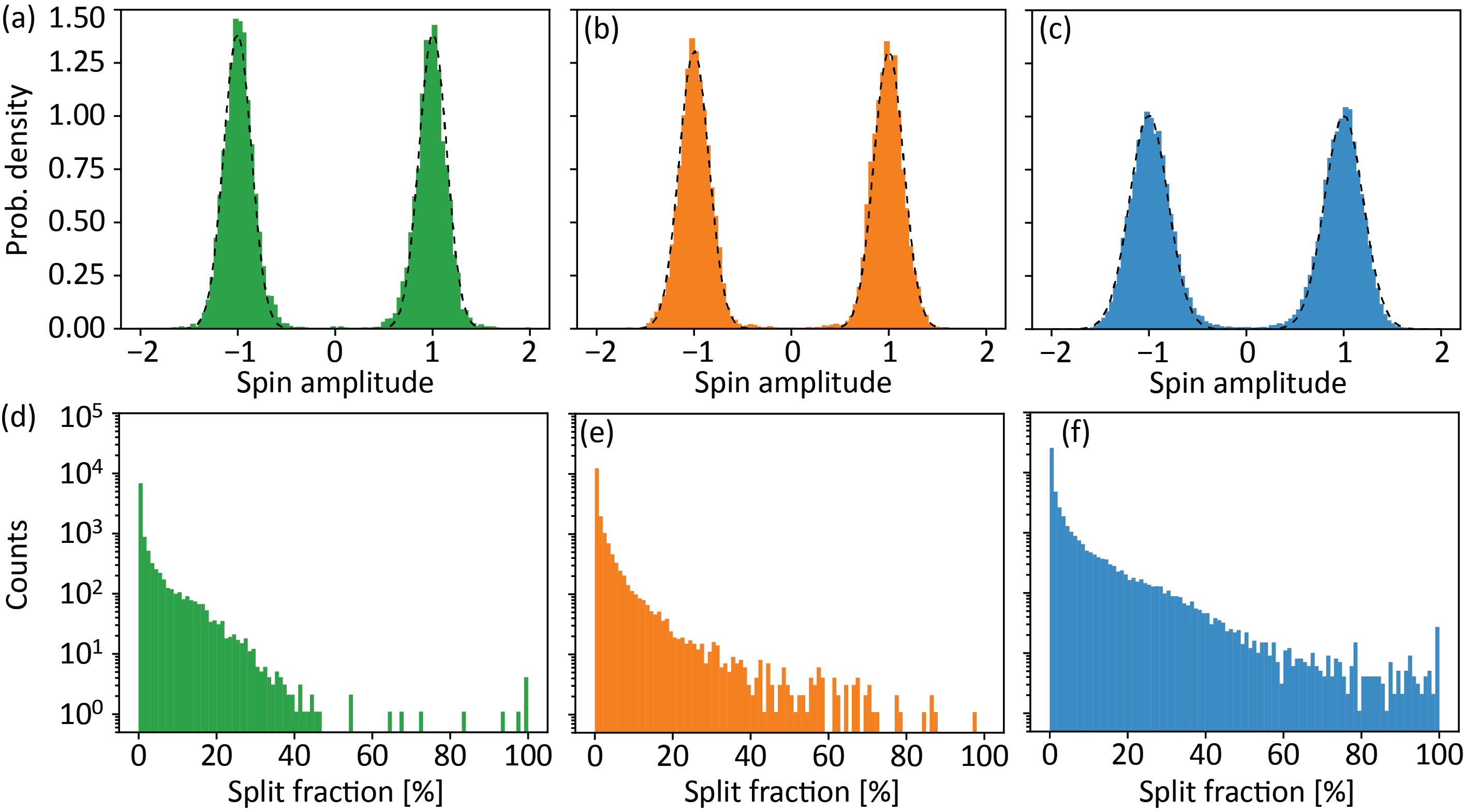}
\caption{ (a) The spin amplitude distribution averaged over all spin sites, replicas, and all disorder realizations for system size $n=8$. Dotted lines show the result of a least-squares fit to a bimodal Gaussian form with standard deviation $\sigma=0.14$. (b) The same for $n=12$ with $\sigma=0.15$ and (c) $n=16$ with $\sigma=0.20$. (d) The distribution of split fractions $(a^i_{01})^2+(a^i_{10})^2$ within the ensemble of data for $n=8$.  It is averaged over all spin sites, replicas, and disorder realizations. The mean split fraction is 2\%; note the log vertical scale. (e) The split fractions for $n=12$ also have a mean of 2\%. (f) The split fractions for $n=16$ have a mean that increases to 4\%.
}
\label{fig:amps}
\end{figure}

We perform a least-squares fit to extract the parameters $\sigma_{x/y}$ and $\{A_i,\mbf{r}_i, a^i_{00}, a^i_{10}, a^i_{01}\}$ for each spin. An example fit result is shown in Fig.~\ref{fig:imageAnalysis}d with the residual of the fit shown in Fig.~\ref{fig:imageAnalysis}e. We achieve typical errors of $5\%$ for fits to $n=16$ spins, defined as the normalized sum of the squared difference between the fitted and cavity emission fields. The low fit error indicates that using $\varphi=0$ is a good approximation in our parameter regime; recall that $\varphi=0$ sets the cavity modes to be perfectly degenerate in the Green's function. Typical fitted widths are $\sigma_x = 5.2(2) \mu$m and $\sigma_y=5.4(2) \mu$m. The extracted spin components are $\ex{\hat{S}_i^x}=A_ia^i_{00}/\mathcal{N}$, where the normalization $\mathcal{N}$ is chosen such that $\sum_{i=1}^n\ex{\hat{S}_i^x}^2=1$. The measured distributions of $\ex{\hat{S}_i^x}$ are shown in Fig.~\ref{fig:amps}.  The components $\text{HG}_{01}$ and $\text{HG}_{10}$ average to zero over the ensemble and so they are excluded from $\ex{\hat{S}_i^x}$.    We  see that the spin amplitudes are approximately binarized, with little weight near zero. The level of binarization seems to decrease with increasing $n$. This is likely due to the relative weakening of the local interactions within atomic ensembles versus the nonlocal interactions between them. Binarization can be enhanced in the future by increasing local interactions. This might be possible with cavities hosting more perfectly degenerate mode families or by using different $M/N$ resonators that support stronger local interactions, such as the $2/3$ cavity.  We also show in Fig.~\ref{fig:amps} that the split components $(a^i_{01})^2 + (a^i_{10})^2$ remain below 5\% on average. Increasing the strength of local interactions may further diminish the split components. 

The spins distribution within each ensemble may be visualized by subtracting the nonlocal components of the $4/7$ field from the measured images. This leaves only the sources of the field, which are the effective spins from the atomic ensembles. This is computed by isolating the nonlocal component of the fitted field; that is found by excluding the source term of the Green's function in Eq.~\eqref{Eq:47Greens}. The nonlocal component is then subtracted from the measured image to provide an estimate of the local effective spin distribution within each ensemble. An example of the local spin distribution image is shown in Fig.~\ref{fig:imageAnalysis}f. This analysis is performed for all images of spin configurations presented in the main text.

\section{Entropy of spin-state ensembles}\label{sec:entropy}

The entropy presented in Figure~\ref{fig:Fig2}b is calculated in the following manner. We perform $n_\text{reps}=200$ trials of the experiment for each ramp time. Each of the 200 trials yields a replica spin configuration $\mbf{s}^\alpha=( \langle \hat{S_1^x} \rangle,\cdots, \langle \hat{S}_{16}^x) \rangle /\mathcal{N}$, where the normalization $\mathcal{N}$ is chosen such that $\mbf{s}^\alpha\cdot\mbf{s}^\alpha=1$ and the replica index $\alpha$ ranges from 1 to $n_\text{reps}$. We now wish to compute the entropy of the set of spin states $\{\mbf{s}^\alpha\}$. In this analysis we retain only the sign of the spin amplitudes to avoid dependence on how the continuous amplitudes are binned. We compute the sample probability distribution
\begin{equation}
    p(\mbf{s}) = \frac{1}{n_\text{reps}}\sum_{\alpha=1}^{n_\text{reps}}\begin{cases}
        1 & \text{sgn}(\mbf{s}^\alpha)=\mbf{s} \\
        0 & \text{sgn}(\mbf{s}^\alpha)\neq \mbf{s}
    \end{cases}, 
\end{equation}
where $\text{sgn}(\mbf{s}^\alpha)=(\text{sgn}(\ex{\hat{S}_1^x}),\cdots,\text{sgn}(\ex{\hat{S}_{16}^x}))$ and $\mbf{s}$ ranges over all $2^{n}$ spin signatures $\mbf{s}\in\{\pm 1,\cdots,\pm 1\}$ for the $n=16$ spins. $p(\mbf{s})$ is symmetrized over the Ising $\mathbb{Z}_2$ symmetry so that $p(\mbf{s})=p(-\mbf{s})$. The sample entropy is then given by
\begin{equation}
    H(\{\mbf{s}^\alpha\})=-\sum_{\mbf{s}}p(\mbf{s})\log_2 p(\mbf{s}).
\end{equation}
The minimal entropy $H(\{\mbf{s}^\alpha\})=1$ is attained in the limit where all $\mbf{s}^\alpha$ have the same spin signature, up to $\mathbb{Z}_2$ symmetry, and the maximal entropy $H(\{\mbf{s}^\alpha\})=n$ is attained in the limit where $p(\mbf{s})=1/2^n$ is a uniform distribution. 

The sample entropy $H(\{\mbf{s}^\alpha\})$ is known to be a biased estimator, typically underestimating the true entropy, while the jackknife estimator yields an unbiased estimate through a resampling procedure~\cite{Zahl1977jai}. This has the form $n_\text{reps}H(\{\mbf{s}^\alpha\}) - (n_\text{reps}-1)H^{(\cdot)}(\{\mbf{s}^\alpha\})$, where $H^{(\cdot)}(\{\mbf{s}^\alpha\})$ is the average sample entropy after leaving out from $\{\mbf{s}^\alpha\}$ a single spin configuration. We use this jackknife method to produce unbiased estimates of the entropy in the main text. 

\section{$J$ matrix analysis}\label{sec:eigs}

The $J$ matrix of a spin glass should be both disordered (in, e.g., the signs of the off-diagonal elements) and induce spin frustration. Frustration occurs when competing interactions make it impossible to minimize all interaction energies. Specifically, this occurs when the product of $J_{ij}$ elements is negative over any closed loop of spins, e.g., when $J_{ij}J_{jk}J_{ki}<0$. We quantify frustration using the fraction of ferromagnetic (FM) --to-- antiferromagnetic (AFM) bonds and by the prominence of frustrated triangles of spins. A measure of disorder can be obtained through an eigenvalue analysis. It is expected that the distribution of eigenvalues approaches a semicircle in the limit of strong disorder, corresponding to the Gaussian orthogonal ensemble (GOE) of random matrices, as considered in the archetypal SK spin glass~\cite{Sherrington1975smo}. In this section, we perform these analyses for the experimentally realized $J$ matrices as described by Eqs.~\eqref{eq:J} and~\eqref{Eq:47Greens}; this study is similar to that which we presented in the theory work of Ref.~\cite{Marsh2021eam}.

The $J$ matrices realized in $4/7$ cavities transition from a ferromagnetic regime to a spin glass regime as the spins are placed farther away from the cavity center. A similar transition is observed for confocal cavities~\cite{Marsh2021eam,Kroeze2023rsb}. This can be seen by considering positions $|\mbf{r}|\ll w_0$ close to the cavity center in Eq.~\eqref{Eq:47Greens} for the Green's function of the $\eta=0$ cavity. In this limit,  all terms in Eq.~\eqref{Eq:47Greens} are positive, realizing only FM couplings. However, for positions $|\mbf{r}|\gtrsim w_0$ farther from the cavity center, the Green's function oscillates in sign, producing both FM and AFM couplings. It is in this regime $|\mbf{r}|\gtrsim w_0$ that the $J$ matrix becomes a spin glass.  

\begin{figure}[t]
    \centering
    \includegraphics[width= 1\linewidth]{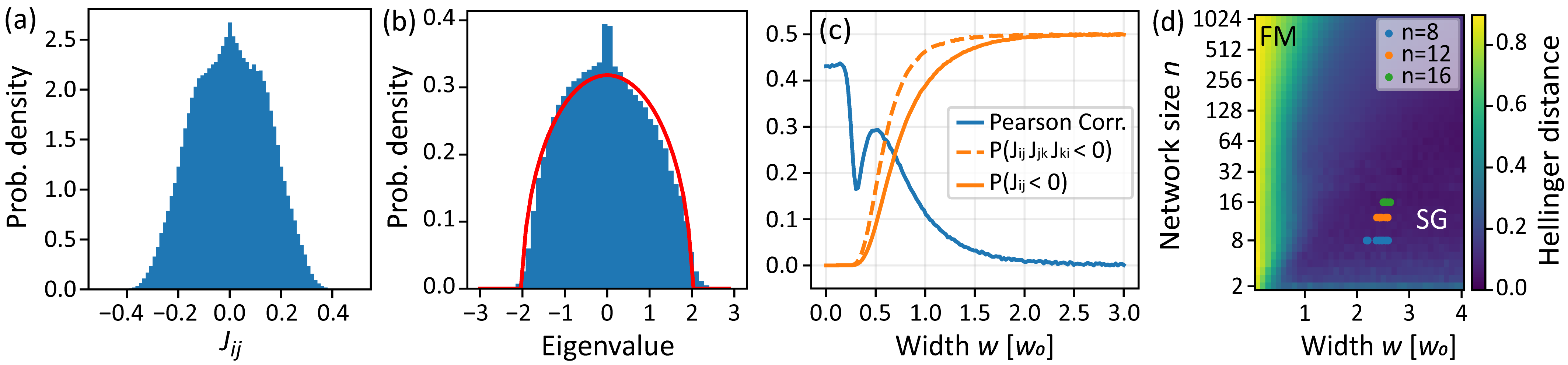}
    \caption{
    (a) Distribution of elements of the $J$ matrix at $w/w_0=2.5$. (b)  Eigenvalue distribution averaged over $2\times10^4$ $J$ matrices for $n=16$ at $w/w_0=2.5$ (blue), shown with a semicircle with radius 2 superimposed (red). The Hellinger distance is 0.08. (c) Correlations among elements of the coupling matrix (blue) decay with increasing  $w$, indicating random coupling. The probability of a frustrated triplet $P(\langle J_{ij} J_{jk} J_{ki} \rangle) < 0 $ and of a single negative coupling element ($P(J_{ij} < 0)$) approach 1/2, further indicating random connectivity and onset of frustration. Error bars are smaller than trace widths.  (d) Hellinger distance between a disorder-averaged $J$ ensemble and the semicircle distribution. All the disorder realizations at $n = 8$,12,16 are plotted in blue, orange and green, resp.}
    \label{fig:HellingerDistance2D}
\end{figure}

We explore the crossover from the FM to spin glass regime as a function of the system size $n$ and the average displacement of the spins from cavity center. We consider spin positions $\mbf{r}_i$ chosen randomly from a 2D Gaussian distribution located at the cavity center and with a tunable standard deviation $w$, as was considered in our analysis of the confocal cavity~\cite{Marsh2021eam}.  For each combination of $n$ and $w$,  we generate 50 disorder realizations of the $J$ matrix and compute the eigenvalue distribution of each. The eigenvalues are normalized by their standard deviation for simplicity. We neglect self-interaction terms in $J$ since these shift the eigenvalue distributions only globally.  We then compute the disorder-averaged eigenvalue distribution $p(\lambda)$ and compare it to the semicircle distribution $p_\text{sc}(\lambda)\propto \sqrt{1-\lambda^2/R^2}$, where $R=2$ is the radius of the normalized semicircle distribution. We compare $p(\lambda)$ and $p_\text{sc}(\lambda)$ using the Hellinger distance $H(p(\lambda),p_\text{sc}(\lambda))$ defined on the interval $[0,1]$. This is given by 
\begin{equation}
H^2(p(\lambda), p_\text{sc}(\lambda)) = \frac{1}{2} \sum_\lambda \left(\sqrt{p(\lambda)} - \sqrt{p_\text{sc}(\lambda)}\right)^2.
\end{equation}
Values of the Hellinger distance close to zero indicate a close match to the semicircle distribution and thus a glassy $J$ matrix with a high degree of disorder. We show in Fig.~\ref{fig:HellingerDistance2D}a how the Hellinger distance varies with respect to $n$ and $w$. Indeed, we observe that the Hellinger distance becomes small for sufficiently large $w$, corresponding to the spin glass regime. The minimum $w$ for entering the spin glass regime increases with $n$, as was observed for the confocal cavity~\cite{Marsh2021eam}. We estimate that all the experimentally realized $J$ matrices fall well within the spin glass regime;  each of these are plotted as a single points on the figure, where the width $w$ assigned to each point is calculated using the average distance of the atomic ensembles from the cavity center.  

We further analyze the level of frustration in the $J$ matrices and the correlations between their elements in Fig.~\ref{fig:HellingerDistance2D}b. We consider the Pearson correlation $\text{Corr}(J_{ij},J_{jk})$, the probability of AFM couplings $p(J_{ij}<0)$, and the probability of forming a frustrated triangle of spins, quantified by $p(J_{ij}J_{jk}J_{ki}<0)$. These quantities vary with $w$ but do not depend on $n$. We observe that the Pearson correlation vanishes for $w/w_0 \gg 1$, indicating that the $J_{ij}$ elements become effectively independent and identically distributed (iid) when the spins are placed far from cavity center. The initial dip is a nonuniversal feature that changes depending on $\eta$ and the center location of the position distribution. The probability of AFM couplings and the probability of generating a frustrated triangle are similar; both increase from zero, starting near $w=0.3w_0$, and plateau at 50\% by about $2w_0$. This shows that the couplings become FM and AFM in equal proportion and that any triple of spins has a 50\% probability to be frustrated, which is similar to the SK spin glass.   

\section{System size $n=25$ spin glass}

\begin{figure}[t]
    \centering
    \includegraphics[width=1\linewidth]{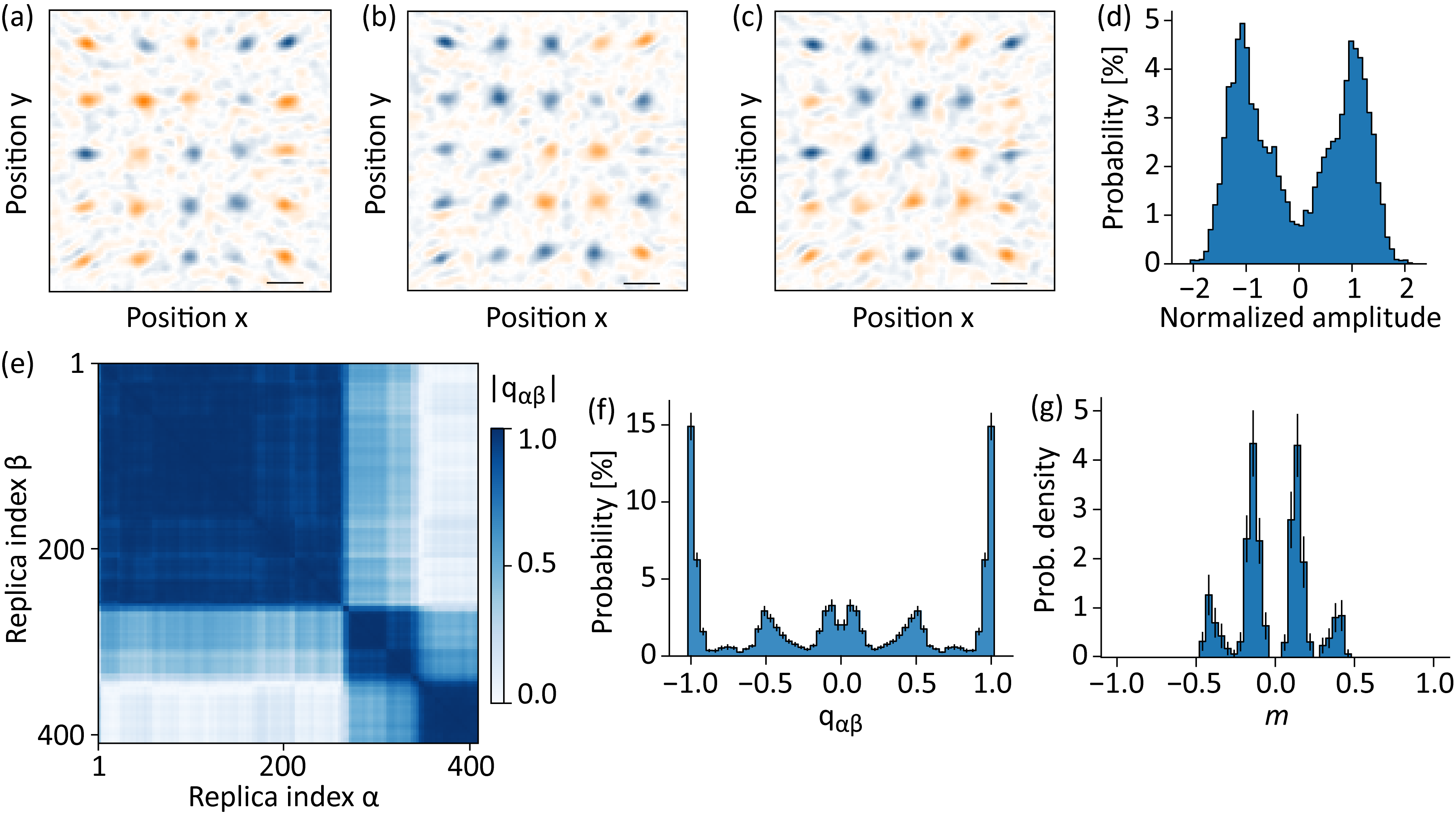}
    \caption{(a-c) Three measured replica spin states of the $n=25$ spin glass. Image in panel (a) is identical to that in Fig.~\ref{fig:Fig1}(a). (d) Spin amplitude distribution measured over 400 replicas. (e) Overlap matrix $q_{\alpha\beta}$ between all replicas. (f) Overlap distribution derived from (e). (g) Magnetization distribution over all replicas.  Error bars come from bootstrap resampling. }
    \label{fig:n25}
\end{figure}

Realizing spin glasses of up to size $n=25$ spins is possible with the current apparatus. We present here the results from a single disorder realization of an $n=25$ spin glass. The number of atoms per spin ensemble is reduced by approximately 15\% compared to the spin ensembles for $n=16$, limited by the maximum number of atoms that can be trapped and cooled in a single experimental run. This, along with the possible contribution of increased frustration at the larger system size, reduces the amount of light scattered from each ensemble.  The critical power $\Omega_c^2$ for the superradiant transition increases in consequence. We compensate for both of these effects by reducing the cavity detuning from $\Delta_C=-2\pi\cdot 20$~MHz to $-2\pi\cdot10$~MHz while holding all other parameters of the ramp schedule the same. Although detuning closer to resonance can modify the cavity Green's function~\cite{Vaidya2018tpa,Kroeze2023hcu}, we verify that our fit residuals remain below $5(0.6)$\%, showing that this effect is minimal. Figure~\ref{fig:n25}(a-c) shows examples of three replica spin states. 

Figure~\ref{fig:n25}d shows the spin amplitude distribution accumulated over 400 measured replica spin states. We find that the spin amplitudes are less binarized than for $n=16$ spin glasses. As mentioned above, the  local interactions are relatively weaker at large $n$ for the 4/7 cavity, and scales as approximately $\sqrt{n}$ in the spin glass phase. A 2/3 cavity would better serve in this regard. 

We now discuss the overlap and magnetization order parameters. Figure~\ref{fig:n25}e shows the full overlap matrix $q_{\alpha\beta}$ between all 400 replicas. Similar to $q_{\alpha\beta}$ found for $n=16$, we observe a nested block diagonal structure. Figure~\ref{fig:n25}f shows the overlap distribution. Goalpost peaks are located at $q_{\alpha\beta}=\pm1$, indicating that the spin glass remains in a deeply organized state. The interior structure demonstrates multiple peaks, as observed for other $n$, indicative of RSB. The magnetization distribution is expected to fluctuate about $m=0$ with a standard deviation of approximately $1/\sqrt{n}=0.2$ for $n=25$. The measured magnetization distribution in Fig.~\ref{fig:n25} is consistent with this prediction and thus does not indicate ferromagnetic order. In conclusion, we find evidence of RSB in a deeply ordered $n=25$ spin glass, similar to that observed for $n=16$ but with increased spin amplitude fluctuations.  

\section{Full set of overlap and magnetization distributions}

\begin{figure}[t]
    \centering
    \includegraphics[width=1.0\linewidth]{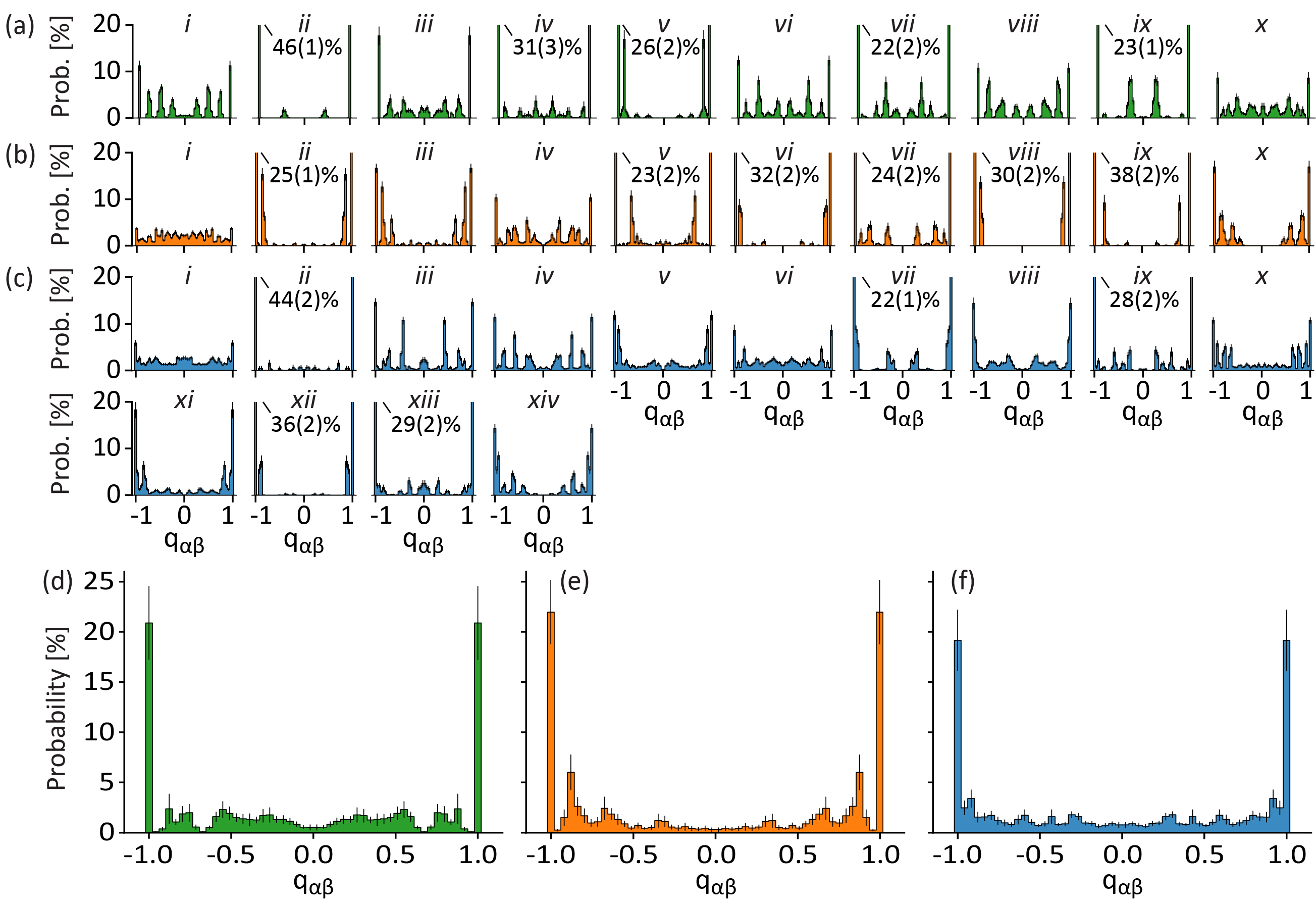}
    \caption{Full set of overlap distributions. The overlap distribution for each disorder realization is presented for (a) $n=8$, (b) $n=12$, and (c) $n=16$. Bins extending beyond the upper limit of the plot are explicitly labeled. Note that (c)\textit{i} corresponds to Fig.~\ref{fig:Fig2}i, column III, in the main text. In addition, (c)\textit{i}-\textit{vii} correspond to Fig.~\ref{fig:Fig3}a \textit{i}-\textit{vii} in the main text.
    The Parisi distributions are presented for (d) $n=8$, (e) $n=12$, and (f) $n=16$. Note that (f) corresponds to Fig.~\ref{fig:Fig3}c in the main text.}
    \label{fig:olaps}
\end{figure}

\begin{figure}[t]
    \centering
    \includegraphics[width=1.0\linewidth]{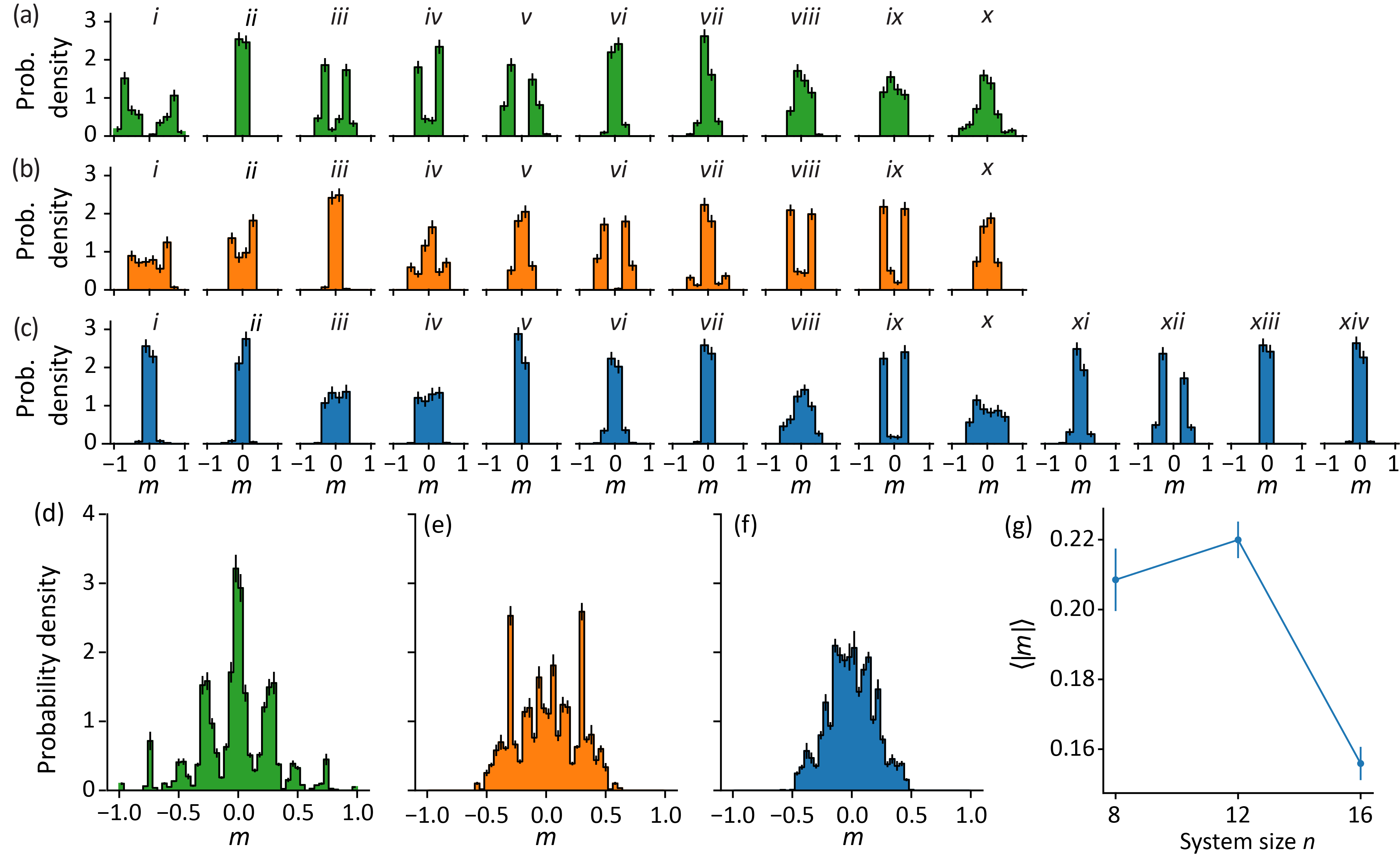}
    \caption{Full set of magnetization distributions for system sizes (a) $n=8$, (b) $n=12$, and (c) $n=16$. The disorder-averaged magnetization distributions are shown for system sizes (d) $n=8$, (e) $n=12$, and (f) $n=16$. (g) The average absolute magnetization versus system size. Error bars are derived from bootstrap resampling of the disorder-averaged magnetization distributions. }
    \label{fig:mags}
\end{figure}

In this section, we present all measured overlap distributions, Parisi distributions, and magnetization distributions for $n=8$, 12, and 16. We increase the number of replicas measured for each $n$ to keep the average error per bin of the overlap distributions to less than $30\%$. Error bars are derived through bootstrap resampling of the replica spin states. Specifically, we measure 200 replicas for each disorder realization of $n=16$, 150 replicas for $n=12$, and 100 for $n=8$. 
Figure~\ref{fig:olaps} shows the full set of overlap and Parisi distributions. Goalpost peaks remain near $\pm 1$ in all measured overlap distributions, while the internal structure of the overlap distribution varies between disorder realizations. Heterogeneity between overlap distributions is an expected feature of spin glasses which persists in the thermodynamic limit of the SK model due to non-self-averaging of the spin glass phase~\cite{Stein2013sga}.  Figure~\ref{fig:mags} shows the full set of magnetization distributions. The magnetization distributions are centered near zero for all system sizes, as expected in a spin glass phase. The absolute magnetization $\ex{|m|}$ is expected to decrease with increasing system size. While $\ex{|m|}$ for $n=8$ and $n=12$ are similar within error bars due to finite sampling of disorder realizations, $\ex{|m|}$ for $n=16$ decreases by approximately 25\%. Taken together, the overlap and magnetization distributions are consistent with a deeply ordered spin glass phase.    

\end{document}